# MÖSSBAUER STUDIES OF COORDINATION COMPOUNDS USING SYNCHROTRON RADIATION

Hauke Paulsen, Volker Schünemann, Alfred X. Trautwein[1], Heiner Winkler

Institut für Physik, Universität zu Lübeck, Ratzeburger Allee 160, D-23538 Lübeck, Germany

**ABSTRACT**

Nuclear resonant forward scattering (NFS) and nuclear inelastic scattering (NIS) of synchrotron radiation are fairly recent spectroscopic methods for the investigation of complexes containing Mössbauer-active transition metal ions. NFS, which can be regarded as Mössbauer spectroscopy in the time domain, overcomes some limitations of conventional Mössbauer spectroscopy as has been demonstrated especially for bioinorganic compounds. NIS extends the energy range of conventional Mössbauer spectroscopy to the range of molecular vibrations. Since NIS is sensitive only to the mean-square displacement of Mössbauer nuclei it can be used as site-selective vibrational spectroscopy. It complements usefully comparable techniques such as IR or Raman spectroscopy. Examples are given for applications to spin crossover complexes, nitroprusside compounds, heme model complexes and myoglobin.

---

[1] Corresponding author. Fax: ++49 451 500 4214, e-mail: trautwein@physik.uni-luebeck.de

**CONTENTS**

**ABBREVIATIONS**

| | |
|---|---|
| APD | avalanche photo diode |
| bpp | 2,6-bis(pyrazol-3-yl)pyridine |
| DFT | density functional theory |
| efg | electric field gradient |
| EPR | electron paramagnetic resonance |
| HS | high spin |
| LS | low spin |
| $MS_{1,2}$ | metastable states 1 and 2 of the nitroprusside anion |
| msd | mean-square displacement |
| NFS | nuclear resonant forward scattering |
| NIS | nuclear inelastic scattering |
| PDOS | partial phonon density of states |

tpa            tris(2-pyridylmethyl)amine)
tptMetame     1,1,1-tris((N-(2-pyridylmethyl)-N-methylamino)methyl)ethane

# 1. INTRODUCTION

Mössbauer spectroscopy is based on the recoilless nuclear γ-resonance discovered by Rudolf Mössbauer in 1958 [1]. A γ-quantum emitted by a Mössbauer nucleus can be absorbed in resonance by another nucleus of the same kind if the recoil energy of the absorbing nucleus is small compared to the linewidth of the nuclear transition ($\Gamma \sim 5 \cdot 10^{-9}$ eV for $^{57}$Fe). This is the case if the recoil is transferred not to the absorbing nucleus alone ($E_R \sim 2$ meV) but to the embedding lattice as a whole ($E_R \sim 0$) and if the same holds for the emission of the γ-quantum. There is a finite probability for such a zero-phonon process that is given by the so called Lamb-Mössbauer factor $f_{LM}$. This factor increases with decreasing temperature. If the emitting and the absorbing nuclei do not have the same chemical environment, the nuclear resonance can be destroyed by small shifts of the nuclear energy levels due to hyperfine interactions between the Mössbauer nucleus and the surrounding nuclei and electrons and external fields. Such energy shifts can be detected by tuning the energy of the γ-quanta, and in this way valuable information can be obtained about the geometric and electronic structure of the molecule or crystal that hosts the Mössbauer nucleus.

In conventional Mössbauer spectroscopy a simple setup consists of a source, a detector for γ-quanta, and a sample that is mounted between source and detector (transmission geometry). The source, which can be moved with a tunable velocity of some mm/s, contains a radioactive precursor that decays to the excited state of the Mössbauer nucleus, which in turn relaxes into the ground state by emission of a γ-quantum (e.g. $^{57}$Co decays to $^{57}$Fe). Resonance can be observed if the velocity of the source is chosen in such a way that the Doppler shift of the γ-quantum compensates for the hyperfine interactions of the Mössbauer nucleus with its surroundings. As a consequence, a fraction of the γ-quanta emitted from the source are absorbed by the sample and the count rate of the detector decreases. In the last several decades this method has developed into a powerful tool in solid-state physics and chemistry, including bioinorganic chemistry [2]. It has been applied to more than 80 so-called Mössbauer-active isotopes such as $^{40}$K, $^{61}$Ni, $^{67}$Zn, and $^{57}$Fe, to mention only those that are important in bioinorganic chemistry. From these $^{57}$Fe is by far the most intensely investigated Mössbauer isotope. This is, among other reasons, due to the availability of a precursor ($^{57}$Co) with a lifetime (270 d) that is very suitable for most experimental purposes, and also due to, in most cases, relatively high Lamb-Mössbauer factors, which mean short measurement times.

The resulting spectra can be used as fingerprints, or they can be rationalized by the spin-Hamiltonian formalism [3] (see also [4,5] for a review). Within this formalism a variety of parameters that reflect the electronic structure of the molecule containing the Mössbauer nucleus are obtained by fitting a simulated spectrum to the experimental one. Among these parameters are the isomer shift, $\delta$, which gives information about spin and oxidation state of the Mössbauer ion, and the quadrupole splitting, $\Delta E_Q$, which describes (together with the asymmetry parameter $\eta$) the anisotropy of the electric field in the vicinity of the Mössbauer nucleus. If the Mössbauer ion is paramagnetic the degeneracy of its spin multiplet is usually lifted even in the absence of an external magnetic field, due to second-order spin-orbit coupling between different electronic states. The parameters that describe the splitting of the spin multiplet and the interactions between the electronic spin, the nuclear spin, and an external magnetic field in the framework of the spin-Hamiltonian formalism are the zero-field splitting $D$, the rhombicity $E/D$, the hyperfine tensor **A** and the tensor **g** [4,5]. These parameters which can be extracted from Mössbauer spectra reflect the character of the electronic ground state and the lowest excited states of the sample compound.

Mössbauer spectroscopy has become an indispensable tool in many fields of physics and chemistry, but there are limitations that cannot be overcome by the conventional technique, which can be regarded as Mössbauer spectroscopy in the energy domain. With the advent of third-generation synchrotron radiation sources the technique of nuclear resonant forward scattering (NFS) of synchrotron radiation has been developed [6, 7, 8, 9, 10] which can be regarded as Mössbauer spectroscopy in the time domain. Another related technique, nuclear inelastic scattering (NIS) of synchrotron radiation [11], can be regarded as an extension of the conventional, energy-resolved Mössbauer spectroscopy (in the range $10^{-9} - 10^{-7}$ eV) to energies of the order of molecular vibrations (in the range $10^{-3} - 10^{-1}$ eV).

The use of synchrotron radiation overcomes some of the limitations of the conventional technique. For instance, NFS allows the direct determination of the Lamb-Mössbauer factor (examples can be found in [12,13]). In addition the high brilliance and the extremely collimated beam lead to a large flux of photons through a very small size of the sample (0.1 – 1 mm$^2$) that make it possible to measure extremely small samples of bioinorganic compounds or metalloproteins. Information about the anisotropy of a single crystal can also be obtained because of the polarization of the synchrotron radiation (a textbook example is given in [12]). The energy of the synchrotron radiation can be tuned over a wide range and therefore it will permit measuring those Mössbauer nuclei for which no suitable conventional Mössbauer source is available. The most promising examples are $^{61}$Ni [14] and $^{67}$Zn where the lifetimes of the precursors, 99 min for $^{61}$Co and 78 h for $^{67}$Ga, are far too short for most experimental purposes. Some of the basic features of NFS and conventional Mössbauer spectroscopy are compared in Table 1.

Synchrotron radiation can be tuned over a wide energy range, whereas in the conventional setup the energy of the γ-quanta can be tuned only in the range of some 100 neV. Conventional Mössbauer spectroscopy therefore gives only limited information (comprised in the Lamb-Mössbauer factor) about the dynamics of the Mössbauer nuclei, while NIS spectra provide detailed spectral information about their dynamics. This method shares some similarity with the resonance Raman method. The major difference is that instead of an electronic resonance, a nuclear resonance is used in NIS. The NIS method is a site-selective method: the intensity of individual peaks in the spectrum is roughly proportional to the mean-square displacement of the Mössbauer nucleus arising from the corresponding molecular vibration. The measured NIS spectra can be directly interpreted in terms of molecular vibrations since no electronic properties such as polarizability or hyperpolarizability influence the intensity of the spectrum.

The following chapter describes both NFS and NIS methods and the application of density functional theory (DFT) to the interpretation of NFS and NIS spectra. The third chapter presents selected applications of these methods in the field of bioinorganic and inorganic chemistry. The fourth chapter deals with specific topics such as relaxation phenomena.

Table 1: Comparison between NFS and conventional Mössbauer spectroscopy.

|  | Resonant forward scattering | Conventional resonant absorption |
|---|---|---|
| Sample cross section | 0.1 – 1.0 mm$^2$ | 10 – 100 mm$^2$ |
| Amount of 57Fe | ≈ 3 μg | ≈ 30 μg |
| Data acquisition time | 1 – 5 h | 1 – 5 d |
| Background | ≤ 1 % | ≥ 90 % |
| Data analysis | complex, only visually controlled simulations possible | simple, preliminary least squares fits with Lorentzians possible |

## 2. METHODS

### 2.1. NUCLEAR RESONANT FORWARD SCATTERING

As mentioned in the Introduction, the primary parameters which can be extracted from conventional Mössbauer spectra are the Lamb-Mössbauer factor as well as the various fine and hyperfine parameters which provide information about the state of the electronic environment of the Mössbauer nuclei.

Nuclear resonant forward scattering (NFS) is in a similar way sensitive to the nuclear environment so that time-differential spectra provide precise values of hyperfine frequencies and relative isomer shifts. In addition, it is an elastic and coherent scattering process, i.e., it takes place without energy transfer to electronic or vibronic states and is delocalized over many nuclei. Due to the temporal and spatial coherence of the radiation field in the sample a characteristic pattern of so-called dynamical beats develops during the propagation through the sample. This pattern can be used to determine the effective thickness and thus the Lamb-Mössbauer factor of the sample in a much better way than in conventional Mössbauer spectroscopy. The time-differential intensity of NFS by a nuclear state of width $\Gamma_0 = \hbar/\tau_0$, which is subject, e. g., to an electric quadrupole splitting $|\Delta E_Q| = \hbar \cdot \Delta\omega$ with $|\Delta E_Q| \gg t_{\text{eff}} \Gamma_0$ can be approximated by [15]

$$I_{\text{fwd}}(t) \sim \frac{\Gamma_0}{\Delta E_\gamma} \cdot \frac{t_{\text{eff}}}{t/\tau_0} \cdot \exp(-t/\tau_0 - \sigma_{\text{el}} nd) \\ \times J_1^2\left(\sqrt{0.5 t_{\text{eff}} t/\tau_0}\right) \cdot \cos^2\left(\tfrac{1}{2}\Delta\omega t + \frac{t_{\text{eff}} \Gamma_0}{8\hbar\Delta\omega}\right) \quad (1)$$

where $\Delta E_\gamma$ is the bandwidth of the incoming radiation and $\sigma_{\text{el}}$ is the electronic absorption cross-section. The exponential decay is obviously modulated by the square of a Bessel function of first order giving rise to the above-mentioned dynamical beats. The positions of their minima and maxima can be determined with high accuracy and thus give precise information about the effective thickness of the sample,

$$t_{\text{eff}} = dn f_{\text{LM}} \sigma_0 , \quad (2)$$

which appears in the argument of $J_1$. Because the geometric thickness $d$ of the sample, the number of Mössbauer nuclei per unit volume $n$, and the nuclear absorption cross-section at resonance $\sigma_0$ are generally known, the Lamb-Mössbauer factor $f_{\text{LM}}$ can be determined with high accuracy without having to correct for background and source properties as in conventional Mössbauer spectroscopy.

In addition to the dynamical beats the so-called quantum beats, described by the cosine term in eq. (1), can be extracted from an NFS spectrum. In principle, these quantum beats are the results of interferences between the radiation scattered by different nuclear resonances. Consequently, their frequencies correspond to energetic differences between these resonances. However, the mathematical description of the time-differential NFS intensity is, in most cases (e.g. in cases when frozen solutions are investigated), not as straightforward as it may appear in eq. (1). The reason is that couplings between the various components of the delocalized radiation field in the sample have to be taken into account by an integration over all frequencies. The problem has been solved in different ways in a series of program packages, the most prominent of which are called CONUSS [16,17], MOTIF [18,19] and SYNFOS [20,21].

## 2.2. NUCLEAR INELASTIC SCATTERING

In contrast to NFS, nuclear inelastic scattering (NIS) takes place with vibrational energy transfer and is localized on a particular nucleus so that it proceeds incoherently. This method allows the nuclei to be excited by radiation that is several meV higher or lower than the resonance energy, because the energy difference between the incident beam and this resonance energy is transmitted to a phonon in the lattice or a vibrational mode of the molecule. In this sense NIS is similar to resonance Raman spectroscopy, except that a nuclear transition, and not an electronic transition is used to excite vibrational modes. NIS is therefore suitable for complementary investigations of molecular vibrations by infrared and Raman spectroscopy because it aims at the Mössbauer nucleus alone and allows for the extraction those normal modes that are connected with a considerable mean-square displacement (msd) of the Mössbauer nucleus. It thus provides information on the partial phonon density of states (PDOS) with an energy resolution of in the order of meV.

The absorption probability, which is registered in an NIS measurement, is a function of the energy difference $E = E_{in} - E_{res}$ between the energy of the incoming radiation $E_{in}$ and the resonance energy of the Mössbauer nucleus $E_{res}$ and depends in general also on the direction of the incoming radiation represented by its wavevector **k**. In the low-temperature approximation this function can be written as

$$S(E,\mathbf{k}) \propto \sum_i \delta(E-E_i)\langle(\mathbf{k}\cdot\mathbf{u}_i)^2\rangle ,\qquad(3)$$

where $E_i$ is the energy of the molecular vibration. The term $\langle(\mathbf{k}\cdot\mathbf{u}_i)^2\rangle$ describes the contribution of the vibration to the total msd of the iron nucleus projected onto the wavevector **k**. $S(E)$ is to be taken as normalized, i.e., by definition $\int S(E)\mathrm{d}E = 1$. Different from vibrational spectra detected with other methods such as Raman spectroscopy, NIS spectra depend solely on the eigenfrequencies and eigenvectors of the vibrations, but not on other properties like the polarizability of the molecule.

In the approximation of eq. (3) the absorption probability reflects directly the PDOS. At elevated temperatures where multiphonon excitations take generally place, a more sophisticated evaluation is required to extract the PDOS from the NIS spectra. The program called PHOENIX, which is also described in ref. [17], performs this task.

## 2.3. INSTRUMENTATION

Nuclear inelastic scattering exploits the unique features of synchrotron radiation, i.e. high brilliance and collimation of the X-ray beam and its pulsed time structure. A typical experimental set-up of this technique as installed at the Nuclear Resonance beamline ID 18 of the European Synchrotron Radiation Facility (ESRF) [22] in Grenoble/ France is shown in Fig. 1. A monochromatic beam of X-rays with about one eV bandwidth is prepared by the standard beamline equipment, among them the undulator and the high-heat-load pre-monochromator being the most important items. Further monochromatization down to ~meV bandwidth is achieved with the high-resolution monochromator.

The beam of X-rays irradiates a sample and excites the resonant nuclei. The radiation that results from nuclear de-excitation has to be distinguished from the photons, that pass through the sample without interaction and from those that are scattered by the electrons. This is achieved by utilizing the time distribution of the scattered radiation. The yield of nuclear scattering is delayed due to the finite lifetime of the nuclear excited state (~100 ns), whereas the electronic scattering is essentially prompt on that time scale. Fast electronic devices are synchronized with the revolutions of the electron beam in the storage ring and count only the events between the pulses of incident radiation. Thus only the delayed quanta which result from nuclear scattering are detected.

Nuclear scattering is counted by two avalanche photo diode (APD) detectors. The first detector (#1 in Fig. 1) is located close to the sample. It counts the quanta scattered in a large solid angle. The second detector (#2) is located far away from the sample. It counts the quanta scattered by the nuclei in the forward direction. These two detectors follow two qualitatively different processes of nuclear scattering. Because it is scattered inelastically, the photon acquires a certain phase shift and, therefore, is no longer coherent with the incident radiation. If the phase shift is random for various nuclei, the scattering is spatially incoherent over the nuclear ensemble, and the scattered photons may be associated with some individual nuclei. The products of de-excitation of the individual nuclei are emitted in a large solid angle as spherical waves (neglecting polarization effects). Thus the first detector monitors the energy spectrum of inelastic excitation by counting the incoherent scattering of radiation by individual nuclei. In addition to incoherent scattering of primary radiation, the first detector may also collect delayed atomic fluorescent radiation resulting from internal conversion. This contribution is dominant, for instance, for the most important Mössbauer isotope of $^{57}$Fe. The two channels have identical dependences on the energy of incident radiation.

With the same setup NFS can also be measured by monitoring the intensity as a function of the delay time by the second detector (#2), which is mounted in the forward direction, beyond the sample.

## 2.4. DENSITY FUNCTIONAL CALCULATIONS

Soon after the development of Mössbauer spectroscopy, electronic structure calculations have been applied to calculation of the Mössbauer parameters. It turned out that bonding effects often play an important role in determining the Mössbauer parameters. Atomic calculations using a crystal field model are, therefore, poor approximations. For this reason molecular orbital calculations on a semi-empirical level were used initially to calculate Mössbauer parameters of heme proteins [23,24]. With the development of density functional theory (DFT), DFT-based programs [25] have become the method of choice for the calculation of Mössbauer parameters (see for instance [26,27] for recent results). In most cases DFT calculations give reasonably accurate results for the isomer shift δ, the quadrupole splitting $\Delta E_Q$, and the asymmetry parameter η. In fact, the combination of DFT calculations and Mössbauer measurements is currently the most accurate method for determining the nuclear quadrupole moment of $^{57}$Fe [28].

DFT methods have also been applied to the simulation of NIS spectra. In the case of molecular crystals, if the interactions between *inter-* and *intra-*molecular modes can be neglected, the molecular part of the anisotropic probability density $S(E,\mathbf{k})$ defined in equation (3) can be obtained from the electronic structure calculated for a free molecule [29]. In the harmonic approximation, which is applied here, for a given molecule containing $N$ atoms there exist *3N-6* normal modes of molecular vibrations that can be with DFT methods. The temperature dependent msd of the Mössbauer nucleus can be obtained according to the relation

$$\left\langle \left(\mathbf{k}\cdot\mathbf{u}_j\right)^2 \right\rangle = \left(\mathbf{k}\cdot\mathbf{b}_j\right)^2 \left(E_R / k^2 E_j\right) \coth\left(E_j / 2k_B T\right). \tag{4}$$

Here $E_R$ denotes the recoil energy of the free Mössbauer nucleus and $\mathbf{b}_j$ is the projection of the $j^{th}$ eigenvector of the dynamical matrix into the 3-dimensional subspace of the iron coordinates. The molecular quantity $\mathbf{b}_j$ corresponds to the phonon polarization vector $\mathbf{e}_{j,\text{Fe}}(\mathbf{q})$ in the solid state. In this approximation the molecular part of the PDOS can be formulated as

$$g_{\text{Fe}}^{(\text{mol})} = \sum_j \delta\left(E - \hbar\omega_j\right)\left(\mathbf{k}\cdot\mathbf{b}_j\right)^2. \tag{5}$$

The molecular PDOS defined by equation (5) comprises only those contributions to the PDOS which are due to optical phonons, whereas the contributions from acoustical phonons are not included.

The DFT calculations for the high-spin (HS) and low-spin (LS) isomers of the [Fe(tptMetame)]$^{2+}$ cation referenced in section 3.1 were performed with the B3LYP method, implemented in the Gaussian98 program system together with the Dunning-Huzinaga all-electron double–$\zeta$ basis set for H, C, and N and the Los Alamos effective core potential plus double–$\zeta$ basis set on Fe [30]. The calculated PDOS was line-broadened by folding it with a normalized Gaussian with 1 meV full linewidth that corresponds to the instrumental resolution. The simulated NIS spectra for the nitroprusside anion referenced in section 3.2 have been calculated using Becke's exchange functional [31] and the correlation functional of Lee, Yang, and Parr [32] (BLYP) together with the 6-311+G(2d,p) basis for H, C, and N and the Wachters-Hay double-$\zeta$ basis for Fe [33].

## 3. APPLICATIONS

### 3.1. SPIN CROSSOVER COMPLEXES

Spin crossover complexes exhibit a thermally driven change of the spin state of the metal center. For many of these complexes the transition from the low-spin (LS) state to the high-spin (HS) state and vice versa can also be induced by irradiation with light, and they are therefore promising materials for optical information storage and display devices [34]. The driving force for the thermally induced spin transition is the increase in entropy between the LS and the HS state, the major part of which arises from the decrease of vibrational frequencies that can be observed when passing from the LS to the HS state. The most prominent frequency shift can be attributed to the iron-ligand bond stretching modes that are directly influenced by the change of the spin state. These modes are exactly the modes that give rise to the dominant inelastic peaks of NIS spectra of iron molecular crystals; this explains the value of the NIS method for the understanding of spin crossover. NIS measurements on spin crossover complexes were initially performed for [Fe(tpa)(NCS)$_2$] and for [Fe(bpp)$_2$] with an energy resolution $\Delta E$ of 6 meV and 1.7 meV, respectively [12,35]. These measurements exhibit a shift of the average iron-ligand bond stretching frequencies, but the low energy resolution has prohibited the separation of single vibrational modes. Recent improvements of high-resolution monochromators ($\Delta E \sim 0.65$ meV) [36] allowed individual vibrational modes to be resolved [37]. NIS measurements for the iron(II) spin crossover complex [Fe(tptMetame)](ClO$_4$)$_2$ revealed the angular dependency of the vibrational modes of a molecular crystal [38]. This complex was especially suitable for this kind of investigation, because all molecules in the unit cell have their threefold symmetry axis oriented parallel to the crystallographic **a**-axis.

NIS spectra were recorded for the LS isomer at 30 K and for the HS isomer at room temperature with the crystallographic **a**-axis parallel and perpendicular to the wavevector **k** of the incoming synchrotron radiation. From the experimental spectra, as well as from the spectra calculated with DFT methods, the partial density of vibrational states (PDOS) was extracted. Comparison of the experimental and the calculated PDOS for the HS and the LS isomer gave an overall agreement for the essential features, with the exception of the acoustic contributions, which are not included in the molecular model. The acoustic part, however, extends only up to 5 meV and is roughly the same for the HS and the LS isomer. With the help of the calculated normal modes, the two predominant peaks at 40 and 46 meV for **k** $\perp$ **a**, as well as the predominant peak at 44 meV and the small peak at 41 meV for **k** $\parallel$ **a** in the PDOS of the LS isomer, could be assigned to iron-ligand bond stretching modes. Considering an ideal FeN$_6$ octahedron, there exist six Fe-N bond stretching modes which transform according to the $A_{1g}$, $E_g$, and $T_{1u}$ irreducible representations of the pointgroup $O_h$. The *gerade* modes $A_{1g}$ and $E_g$ are not connected with a displacement of the iron nucleus (i.e. they exhibit zero msd of the iron ion), and thus only the three $T_{1u}$ bond-stretching modes can be detected by NIS. Actually, the first coordination sphere of the iron in the [Fe(tptMetame)]$^{2+}$ complex has to be regarded as a distorted [FeN$_6$] octahedron with $C_3$ symmetry. Due to the lower symmetry the vibrational $T_{1u}$ term is split into an $A$ and an doubly degenerate $E$ term. In this case all six bond-stretching modes are visible by NIS: two pairs of doubly degenerate $E$ modes are connected with a displacement of the iron nucleus within the equatorial plane and are visible if **k** lies within this plane, which is perpendicular to the threefold symmetry axis (Fig. 2). The remaining two $A$ modes are connected with a displacement of the iron nucleus along the molecular symmetry axis (Fig. 2). The smaller peak at 41 meV corresponds to the nearly fully symmetric breathing mode where the iron nucleus participates only slightly in the vibration. The experimental PDOS for the HS isomer at room temperature with **k** $\perp$ **a** (Fig. 3) exhibits one dominant peak at about 30 meV, whereas the calculation yields two peaks at 28 and 33 meV which correspond to iron-ligand bond stretching modes. Apparently these modes are shifted by 10 meV to lower energy when passing from the LS to the HS state.

### 3.2. NITROPRUSSIDES

Sodium nitroprusside dihydrate ($Na_2[Fe(CN)_5NO]\cdot 2H_2O$) as well as many other nitroprusside salts have become a promising basis for holographic information storage devices with extremely high capacity [39] since in 1977 the existence of long-lived metastable states, $MS_1$ and $MS_2$, has been discovered by Mössbauer spectroscopy [40]. But despite all efforts the nature of the metastable states remained in the dark for almost 20 years until Carducci *et al.* [41] were able to characterize the structure of $MS_1$ and $MS_2$ by X-ray diffraction. According to their results $MS_1$ corresponds to an isonitrosyl geometry of the nitroprusside anion and $MS_2$ corresponds to a side-on bonding of the nitrosyl group (Fig. 4). Various electronic structure calculations applying DFT methods [42,43,44,45] confirmed the isonitrosyl structure of $MS_1$, whereas with neutron scattering experiments a nitrosyl structure was found for both the ground state and $MS_1$ [46,47]. To resolve this contradiction NIS turned out to be a complementary technique that could provide evidence for one of the two structures that were proposed for $MS_1$.

NIS spectra for different cuts of a guanidinium nitroprusside (($CN_3H_6)_2[Fe(CN)_5NO]$) monocrystal in the ground state were recorded and interpreted with the help of DFT calculations [29,45]. Additional measurements were performed at 77 K for a sample that was illuminated by blue LEDs (wavelength 450 nm) in order to reach at least partial population of the metastable states (mostly $MS_1$). In a guanidinium nitroprusside monocrystal the symmetry axes of all nitroprusside anions are approximately parallel. This allowed to be gained information about the anisotropy of the molecular vibrations [48]. For the comparison of measured and simulated NIS spectra, the normalized probability density of absorption $S(E,\mathbf{k})$ defined in equation (3) has been plotted (Fig. 5). The measured NIS spectra with and without population of the metastable states were practically identical except for a peak at about 540 $cm^{-1}$. This peak appears in the spectrum that was recorded after illuminating the sample with blue light, and it is missing in the other spectrum before. The simulated NIS spectrum for the ground state (Fig. 5, dashed line) using the normal modes calculated with BLYP/6-311+G(2d,p), is in qualitative agreement with the measured ground state spectrum. The simulated spectrum for the isonitrosyl structure (Fig. 5, dotted line) differs from the simulated ground state spectrum in exhibiting its strongest peak at 564 $cm^{-1}$ ($MS_1$) compared to 667 $cm^{-1}$ in the ground state.

If the wavevector $\mathbf{k}$ of the incident synchrotron radiation is parallel to the crystallographic $c$-axis, only 3 $A_1$ modes (referring to an approximate $C_{4v}$ symmetry of the nitroprusside anion) can be observed in the NIS spectrum: an Fe-N bond-stretching mode at 662 $cm^{-1}$ (mode 1), an Fe-C-N bending mode at 447 $cm^{-1}$ (mode 2), and an Fe-$C_{ax}$ bond-stretching mode at 371 $cm^{-1}$ (mode 3). By comparing the measured intensities of the peaks that are attributed to the Fe-X (X=N,O) stretching mode in the ground state and in the metastable state $MS_1$ with the simulated peaks, about 7 % population of $MS_1$ could be estimated. The combined NIS and DFT studies revealed that only the Fe-(NO) stretching mode (mode 1) exhibits a significant shift if the nitroprusside anion is transferred from the ground state to the metastable state $MS_1$. In summary, the comparison of experimental NIS spectra, which were recorded under conditions excluding and including the metastable $MS_1$ state, with simulated NIS spectra using DFT, support the isonitrosyl structure of the metastable $MS_1$ state as proposed by Carducci et al. [41].

## 3.3. NUCLEAR FORWARD SCATTERING ON HEME MODEL COMPLEXES

Heme models have been studied by NFS both in the oxidized and reduced state. All these studies carried out so far have been performed with highly enriched $^{57}$Fe. The use of model complexes in order to test a new technique such as NFS, with respect to its applicability to biological samples is obvious: model complexes can be prepared in high quantities and can be handled with ease. Thus they not only serve as models for proteins, but also as well understood standards for testing new spectroscopic methods.

The "picket-fence" porphyrin $^{57}FeO_2(SC_6HF_4)(TP_{piv}P)]^-$ [49], which mimics the heme site of cytochrome P450 with a thiolate axial ligand to iron in its oxygenated state is among the first model complexes, that were studied by NFS. This complex is diamagnetic and exhibits an isomer shift of $\delta$=0.33 mms$^{-1}$ and a quadrupole splitting of $\Delta E_Q$ = 2.17 mms$^{-1}$ at 4.2 K in a conventional Mössbauer spectrum [50]. The presence of an electric field gradient (efg) manifests itself in a beat structure of the NFS time domain spectrum (Fig. 6a). This efg-dependent beat structure is superimposed by an effective thickness-dependent dynamic beat structure; both are temperature dependent (Fig. 6b,c) and therefore require specific consideration during data analysis [16].

The ferrous state of the prosthetic group termed P460 of the multi-heme enzyme hydroxylamine oxidoreductase from *Nitrosomonas europeae* has been modelled by the "picket-fence" porphyrin acetate complex, $^{57}Fe(CH_3COO)(TP_{piv}P)]^-$ [51], which was also studied by temperature- and field-dependent NFS. Since the iron is in the paramagnetic ferrous high-spin state the application of magnetic fields induces magnetic hyperfine interaction which is represented by a complex beat structure in the NFS spectrum. The corresponding conventional magnetic Mössbauer spectra of iron proteins are analyzed routinely by the spin-Hamiltonian formalism [4]. In order to apply this formalism to magnetic

Mössbauer spectra in the time domain it was implemented into the software package SYNFOS [20]. Fig. 7 shows the measured and simulated NFS spectra of $^{57}Fe(CH_3COO)(TP_{piv}P)]^-$.

The solid lines in Fig. 7 are simulations with the SYNFOS programm using $S = 2$ and zero-field splitting $D = -0.8$ cm$^{-1}$, rhombicity parameter $E/D = 0$, magnetic hyperfine coupling tensor $A/g_n\mu_n = (-17,-17,-12)$ T, quadrupole splitting $\Delta E_Q = 4.25$ mms$^{-1}$, asymmetry parameter $\eta=0$ and effective thickness $t_{eff} = 20$. These parameters have been obtained by conventional Mössbauer spectroscopy [51] and were used as test for the NFS study. This test shows that the spin-Hamiltonian parameters as obtained from conventional Mössbauer studies are useful input parameters for a NFS study which in turn provides a refined parameter set if NFS data extend to high delay time.

An additional test case in this respect is the low–spin ferri-heme complex bis(3-aminopyrazole)tetraphenylporphyrinatoiron(III) chloride ([TPPFe(NH$_2$PzH)$_2$]Cl) [52,53] (Fig. 8). This model complex is of interest because it exhibits conventional Mössbauer spectra [53] which are very similar to those of cytrochrome P450$_{cam}$ from *Pseudomonas putida* [54] and to chloroperoxidase from *Caldariomyces fumago* [55]. In Fig. 8 and Table 2 we compare spectra and parameters as obtained from NFS (Fig. 8a) and conventional Mössbauer studies (Fig. 8b).

In conclusion the NFS studies mentioned above show that this technique yields information about fine and hyperfine parameters which have the same quality as those obtained by conventional Mössbauer spectroscopy. For conventional Mössbauer spectroscopy, however, sample volumes > 100µl are required. For NFS, the sample volume is determined by the diameter of the monochromatized synchrotron beam which at nowadays 3$^{rd}$ generation synchrotron sources can already be focussed into a spot of a few micrometers. The development of 4$^{rth}$ generation synchrotron sources together with even more sophisticated beam line optics may allow to obtain NFS spectra of biological samples with volumes of only a few microliters. The above mentioned studies have put the ground for these exciting experiments yet to come.

Table 2: **Spin – Hamiltonian and Mössbauer parameters obtained from simulations of NFS and conventional Mössbauer spectra of ([TPPFe(NH$_2$PzH)$_2$]Cl (Figure 8; taken from [52]).**

| g – tensor [a] | $\Delta E_Q$ [mm/s] | $\eta$ | $A/g_N\mu_N$ [T] |
|---|---|---|---|
| | **NFS spectra** | | |
| $g_{xx}$ = 1.87 $g_{yy}$ = 2.28 $g_{zz}$ = 2.39 | +2.56[b] | from -2.2 to -3.8 [c] | $A_{xx}$ = -46.9 ± 2.0 $A_{yy}$ = 9.5 ± 2.0 $A_{zz}$ = 17.6 ± 1.5 |
| | **conventional Mössbauer spectra** | | |
| | +2.56 | from -1.7 to -3.5 [c] | $A_{xx}$ = -45.6 ± 3.5 $A_{yy}$ = 6.0 ± 5.5 $A_{zz}$ = 16.9 ± 2.5 |

a) derived from EPR [53]

b) sign of quadrupole splitting does not influence the NFS spectra

c) here η is given in the molecular frame with the z – axis perpendicular to the heme plane. However, the coordinate system of the efg tensor is commonly chosen in such a way that $|V_{zz}| \geq |V_{yy}| \geq |V_{xx}|$; in order to reach this convention in the present case the coordinate system of the efg tensor has to be rotated around the y axis of the molecular frame by the Euler angle $\beta$ = 90° and around the new z´ axis by $\gamma$ = 90°. In this new coordinate system $\eta'$ = 0.0 and $\Delta E_Q'$ = -2.56 mm/s.

### 3.4. NUCLEAR INELASTIC SCATTERING ON MODEL COMPLEXES

The NIS method was used to study the vibronic properties of model hemes. The ferrous nitrosyl tetraphenylporphyrin, Fe(TPP)(NO), was among the first heme models to be studied by NIS [56]. The vibrational spectrum of the iron center could be understood in detail with a normal mode analysis based on conventional force fields.

In order to analyze the NIS spectrum of Fe(TPP)NO Sage et al. introduced the mode composition factor $e_{Fe\alpha}^2$ which is the fraction of the kinetic energy of mode α associated with the motion of the iron atom [56]. The mode composition factor $e_{Fe\alpha}^2$ is a measure of the atomic fluctuations of the iron atom, it is proportional to the contribution of mode α to the msd of the iron atom. The excitation probability of Fe(TPP)(NO) derived from the experimental data by normalization to the recoilless absorption line is shown in Fig. 9, together with the mode composition factors obtained from the normal mode analysis. The correct prediction of the positions of the resonance lines is intriguing. However, in this study a heme doming mode, which involves the motion of the iron perpendicular to the heme plane and is therefore of importance for biochemical reactions involving heme proteins, could not been identified.

In a NIS study of single crystals of the deoxy heme-protein model $Fe^{II}(TPP)(2\text{-MeImH})$ a mode has recently been detected at 78 cm$^{-1}$ that shows significant doming of the porphyrin core [57] . An even larger iron doming motion at 25 cm$^{-1}$ has been explained by the coupling of the phenyls of the porphyrin and the imidazole ligand. In the study of Rai et al. [57] crystals of $[Fe^{II}(TPP)(2\text{-MeImH})]1.5C_6H_5Cl$ were investigated. This molecule crystallizes in the triclinic space group with two molecules per unit cell which are related by inversion symmetry. Thus there is only one orientation of the porphyrin plane within the single crystal. The investigation of single crystals is essential for distinguishing between heme-in-plane and out-of-plane modes. If the crystal is mounted with the porphyrin plane parallel to the Synchrotron beam, only in-plane motions are excited, perpendicular mounting, on the other hand, allows the excitation of out-of-plane modes. Fig. 10 shows the data of the $Fe^{II}(TPP)(2\text{-MeImH})$ crystal. The out-of-plane data show resonances at 67, 79 and 110 cm$^{-1}$, the first two being attributed to heme doming modes and the latter to a iron–imidazole streching mode. These vibrations were expected to be absent in the in-plane data, but this is not the case, indicating that the analysis of low-energy vibrations is not straightforward and has to be taken with care.

It should be noted that the spectra shown in Fig. 10 represent the partial phonon density of vibrational states (PDOS), which were obtained from the experimental data by a formalism developed in [58,59] (see section 2.2). The PDOS is a property of the iron center independent of the spectroscopic method used.

## 3.5. NUCLEAR FORWARD SCATTERING ON MYOGLOBIN

NFS studies of iron proteins are limited up to now to the oxygen storage protein myglobin, because of the ease of working with this protein. These studies set the basis for future studies of other proteins. Keppler et al. [60] reported NFS spectra of deoxy-myoglobin in the temperature range from 50 K to 243 K (Fig. 11). The authors could show that the temperature-dependent beat pattern can consistently be described by the Brownian oscillator model [61], which had been successfully applied to analyze the temperature-dependent spectra obtained by the conventional Mössbauer method. Herta et al. observed dynamic structural disorder in oxy-myoglobin by NFS [62]. In this work it was shown that transitions between two conformational substates of the Fe-O-O moiety take place (see section 4.2).

### 3.6. NUCLEAR INELASTIC SCATTERING ON MYOGLOBIN

NIS studies of myoglobin provided a measure of the total msd $<x_{Fe}>^2$ of the heme iron and also its individual contributions at each frequency [56]. The first inelastic X-ray scattering experiments on myoglobin were performed by Achterhold et al. [63]. **They used unenriched protein and a method which is different from what is now called NIS: The energy in the vicinity of the first nuclear transition of $^{57}$Fe (14.4 keV) is varied. The so monochromatized radiation is scattered not only by the iron nuclei of the myoglobin sample, but by all nuclei present in the sample. The energy of the scattered radiation is analyzed by the Mössbauer effect in the following way: The fraction of radiation, which has -after inelastic scattering - the energy to excite the $^{57}$Fe nuclear Mössbauer level, is absorbed by a $^{57}$Fe foil behind the sample. From this analysator foil the Mössbauer radiation is reemitted with a time delay in the ns range and registered by an avalanche photodiode. The rest of the radiation passes the analysator foil without delay and is not registered. Thus only quanta with exactly the energy of the Mössbauer transition (14.4 keV) are registered. This method yields all vibrations of the protein and is not restricted to the iron center itself and thus provides similar information as obtained by an inelastic neutron scattering experiment.**

Sage et al. reported the complete vibrational spectrum of the iron site of deoxy- and CO-myoglobin [64]. They identified for the first time Fe-N(pyrrole) (Fe-$N_{pyr}$) stretching frequencies. A broad feature near 250 cm$^{-1}$ dominates the high-frequency region of the deoxy-myoglobin spectrum obtained with 0.85 meV (6.9 cm$^{-1}$) resolution (Fig. 12). A similiar feature was also observed but not assigned by Achterhold et al. with 4.4 meV (35.5 cm$^{-1}$) resolution [63]. The spectrum of photolyzed CO-myoglobin (Mb*) resembles that of deoxy-myoglobin. Because of high resolution and good statistics the spectrum could be decomposed into four iron vibrations: a Fe-$N_{His}$ stretch was found at 234 cm$^{-1}$, a Fe-$N_{Pyr}$ stretch at 251 cm$^{-1}$ and 267 cm$^{-1}$, as well as a resonance at 285 cm$^{-1}$ which has not been assigned. For CO-myoglobin a Fe-CO stretch at 502 cm$^{-1}$, a Fe-CO bend at 572 cm$^{-1}$, and in addition a continuum up to 400 cm$^{-1}$ was observed. Raising the temperature to 50 and 110 K led to a broad resonance at around 25 cm$^{-1}$. This band has been assigned to translational vibrations of the whole heme moiety, as predicted theoretically [65]; it is believed to be caused by torsional oscillations of the polypeptide backbone and side chains.

The PDOS of the iron of deoxy- and CO-myoglobin and of myoglobin with different degrees of water content was recently determined by Achterhold et al. [66,67]. They verified that the modes with an energy larger 3 meV (24 cm$^{-1}$) are harmonic at physiologically relevant temperatures. Below 3 meV a temperature dependence of the density of states was observed which was interpreted as mode softening in the low-energy regime at physiological temperatures.

The vibrational properties of the iron in met-myoglobin has also been studied by NIS on single crystals of the protein [68]. This procedure has the advantage that vibrational modes can be assigned by the projection of the vibrational amplitude onto the direction of the synchrotron beam. Modes between 4 and 5 meV (32 and 40 cm$^{-1}$) were identified as heme sliding motions; modes with vibrations between 30.2 and 36.5 meV (244 and 294 cm$^{-1}$) (are mainly within the heme plane, and those between 19 and 25.6 meV (153 and 206 cm$^{-1}$) are perpendicular to the heme plane. Fig. 13 shows the NIS spectra of a met-myoglobin single crystal as a function of the rotation angle $\phi$ around the crystallographic *b*-axis. From these studies we can learn about the low-frequency dynamics of iron in heme proteins. **This understanding will help to interpret future NIS studies on heme enzymes and possibly even on their reaction intermediates.**

## 4. SPECIFIC TOPICS

For the understanding of the effects of time-dependent hyperfine interactions on the NFS spectra it is useful to be remember that the excitation energy is delocalized over a large ensemble of nuclei in the system under consideration. This collective nuclear excited state has been appropriately denoted nuclear exciton by Kagan [69]. The life time of such an exciton is of the order of the natural life time of the nuclei. Each process that perturbs the existing coherence within the nuclear exciton enhances its decay. Time-dependent phenomena such as paramagnetic relaxation or dynamic disorder are perturbations of this kind. Since these processes cause fluctuations of the hyperfine interactions they have also an effect on the quantum beat pattern if their rates are comparable with or larger than the hyperfine frequencies. The reason is that the only way by which the nucleus senses the hyperfine interaction is by the precession around the respective quantization axis. This axis is provided either by the direction of the magnetic hyperfine field $B_{hf}$ or by the largest component $V_{zz}$ of the efg tensor. If this direction changes once or several times during a precession period the nucleus can either reorient (intermediate relaxation) or the interaction becomes ineffective and the respective quantum beats disappear completely (fast relaxation) because at very high rates the nucleus cannot follow the changes any longer.

## 4.1. PARAMAGNETIC RELAXATION

A paramagnetic system is, e.g., a single iron containing molecule with unpaired electrons that yield to a total spin $S$. The $(2S+1)$-fold paramagnetic degeneracy can be lifted by the zero-field and Zeeman interactions appearing in the spin-Hamiltonian approximation. Then an internal magnetic field is induced at the position of the nuclei caused by the spin expectation values $\mathbf{s}_\ell = \langle \phi_\ell^{(el)} | \hat{\mathbf{S}} | \phi_\ell^{(el)} \rangle$ of the electronic spin operator $\hat{\mathbf{S}}$ in the $\ell$-th spin subsystem; thus the hyperfine splitting of the nuclear levels may be different even for equivalent sites and for equivalent orientations of the paramagnetic complexes.

So far the coupling of the electronic spin to the 'heat bath'-type subsystems of the sample (phonons, spin subsystem etc.) has been neglected as the paramagnetic relaxation, caused by such couplings, has been taken to be slow. In the opposite case of very frequent random transitions between electronic spin substates, i.e. fast paramagnetic relaxation, these transitions give rise to a stationary internal magnetic field which is determined by the thermal average of the spin vector $\overline{\mathbf{s}} = \sum_\ell w_\ell \mathbf{s}_\ell$ with

$$w_\ell = \exp[-\varepsilon_\ell/(k_B T)] \left\{ \sum_{\ell'} \exp[-\varepsilon_{\ell'}/(k_B T)] \right\}^{-1} ; \qquad (6)$$

$\varepsilon_\ell$ is the energy of the $\ell$-th spin substate, $T$ is the temperature of the sample and $k_B$ is the Boltzmann constant. Both cases, i.e. slow and fast relaxation, can be treated in conventional manner by diagonalization of the electronic spin-Hamiltonian matrix.

However, in the intermediate relaxation cases (i.e. for moderate transition rates, comparable with the hyperfine splitting of the nuclear levels) such a procedure fails and the fluctuations of electronic spin $\mathbf{s}_\ell$ have to be taken into account. Treating the resulting time dependence $\mathbf{s}_\ell^{rel}(t)$ as a sequence of stochastic Markovian jumps of the electronic spin $\mathbf{S}$ between the spin substates has proven a successful model for the description of paramagnetic relaxations [70]. For spin-lattice relaxation the probabilities $w_{\ell\ell'}$ of the transition jumps $\ell' \to \ell$ ($\varepsilon_\ell \leq \varepsilon_{\ell'}$) can be assumed as follows:

$$w_{\ell'\ell} = w_0 \frac{[(\varepsilon_{\ell'} - \varepsilon_{\ell})/k_B]^p}{\exp[(\varepsilon_{\ell'} - \varepsilon_{\ell})/(k_B T)] - 1} \quad . \tag{7}$$

In equation (7) two fitting parameters appear:

1. The constant $w_0$ which determines the strength of spin-phonon coupling. It depends on the electronic transition matrix elements, phonon frequency distribution parameters etc., but its dependence on the spin substate indices $\ell', \ell$ as well as on the temperature of the sample are neglected in our approximation.

2. The power index $p$ which equals exactly 3 in the Debye approximation. Treating $p$ as a variable can compensate to some extent for the imperfection of the coarse continuum model [71].

The $w_{\ell\ell'}$ have to satisfy the detailed balance condition:

$$w_{\ell\ell'} = \exp[(\varepsilon_{\ell'} - \varepsilon_{\ell})/(k_B T)] \, w_{\ell'\ell} \tag{8}$$

The necessary modifications have been implemented in the SYNFOS routines as described in [20,72].

The 'picket-fence' porphyrin complex [Fe(CH$_3$COO)(TP$_{piv}$P)]$^-$ has already been introduced in section 3.3 as a high-spin ferrous complex that exhibits spin-lattice relaxation. This has been analyzed by the conventional Mössbauer spectra obtained at various temperatures in a field of 4 T applied perpendicular to the γ-ray (Fig. 14). The corresponding time-dependent NFS spectra, taken with the setup available at beam line BW4 of HASYLAB in Hamburg, from a sample with effective thickness $t_{eff}$ = 20 are shown in Fig. 15. The external field of 4 T was applied perpendicular to the wavevector **k** and to the polarization (electric field vector) σ of the incoming beam.

The fits in Fig. 15 indicated by the full lines were performed for all temperatures by means of one single value of $3.65 \cdot 10^5$ s$^{-1}$K$^{-3}$ for the scaling parameter $w_0$ in equation (7). The magnetic hyperfine coupling tensor had to be readjusted slightly compared to the values quoted in section 3.3. In addition, the angular integration required for the microcrystalline samples used in the measurement showed that a certain degree of non-random distribution (texture) had to be taken into account. This was done by excluding orientations with $\theta \leq 40°$ from the powder averaging in all fits, with $\theta$ denoting the angle between the applied field and the molecular $z$ axis. The deviations between theory and experiment below 30 ns are due to spurious radiation.

## 4.2. DYNAMIC STRUCTURAL DISORDER

Three different models for the $FeO_2$ moiety in oxy-myoglobin have been suggested: the bent end-on model of Pauling, the $\eta^2$-bound (side-on) model of Griffith and the bent end-on superoxide model of Weiss (Fig. 16). The Griffith model had been discarded in the course of time due to the results of X-ray diffraction studies. The distinguishing feature of the two other models is that the iron is either Fe(II) or alternatively Fe(III). As it seems more probable that the iron would remain in the ligated state still as iron(II) in the low-spin ($S = 0$) state the Pauling model has been mostly favored. Mössbauer spectroscopy had been expected to resolve the problem but the spectra opened more questions than they answered and the precise nature of the $FeO_2$ bond has remained a matter of controversy for a long period of time.

What has to be explained by any model is first of all the rather large quadrupole splitting for the ferrous low-spin ($S = 0$) state [73]. In this state all three $t_{2g}$ orbitals would be doubly occupied and, hence, the electron shell should be almost spherical and the components of the electric field gradient (efg) tensor small. Furthermore, it has been observed that oxy-myoglobin exhibits a strong temperature dependence of the quadrupole splitting, as does deoxy-myoglobin. For the latter an explanation can be given in terms of a Boltzmann population of excited orbital states. In oxy-myoglobin, however, the electron shell ought to be very stable so that excited states are far away in terms of thermal energy. Extensive MO calculations by Herman and Loew [74] showed for the bent end-on model that an asymmetric delocalization of 3d-electron charge due to covalency is present so that a formal $Fe^{III}$-$O_2^-$ description seemed appropriate and the large quadrupole splitting and even the temperature dependence could be explained in this way. However, for molecular orbitals only the total spin is a good quantum number and this was found to be zero. If this spin is artificially separated in a spin 1/2 at the iron coupled antiferromagnetically to another spin 1/2 at the $O_2$-ligand the resulting state lies according to Herman and Loew more than 13,000 $cm^{-1}$ above the ground state. So it is a matter of taste whether one prefers to speak of a $Fe^{II}$-$O_2$ or a $Fe^{III}$-$O_2^-$ moiety in the oxygenated myoglobins.

What has to be explained in addition is that the two lines broaden asymmetrically at intermediate temperatures. For the 'picket-fence' porphyrin, where this behavior is very pronounced, a satisfactory explanation was given by Spartalian et al. [75] by invoking a dynamic process of the kind that the terminal oxygen jumps between four different positions, two and two of which are equivalent with respect to the efg tensor they produce. The spectra could be fitted very well by this model. A corresponding explanation for myoglobin seemed to be impossible because the X-ray structure of oxy-myoglobin indicated that a rotation of the oxygen molecule in the heme pocket is sterically hindered. However, after refining the X-ray structure Phillips [76] suggested that a structural disorder of the oxygen might be possible with a secondary position of the terminal oxygen rotated by about 40°. Nevertheless, no rigorous mathematical treatment of the protein data has been performed since.

After the advent of nuclear resonant forward scattering (NFS) as an additional tool for the study of hyperfine interactions. This problem was re-investigated, because NFS is much less sensitive to external disturbances like mechanical vibrations that can broaden Mössbauer resonance lines. The NFS data show, apart from an accelerated decay of the delayed counts, no striking variations with temperature (Fig. 17). If the spectra are fitted with quadrupole splitting and effective thickness as free parameters using the program package SYNFOS, one obtains a decrease of the splitting but at the same time an increase of the effective thickness, which is physically unreasonable and obviously due to the faster decay with time. We note that the same procedure applied to deoxy-myoglobin yields a decrease of the splitting and at the same time a decrease of the effective thickness, which is physically reasonable [62]. In the present case, the faster decay is obviously caused by the dephasing, which results from a dynamic structural disorder. Since it was known from single crystal studies [77] that the largest component $V_{zz}$ of the efg tensor lies within the heme plane, a model has been constructed which consists of two electric field gradients with both efg components $V_{zz,1}$ and $V_{zz,2}$ lying in this plane but separated by an unknown angle of rotation $\beta$ around the heme normal [62]. The dynamics is introduced by a jump frequency and the Boltzmann population of the secondary position is determined by the energy difference $\Delta E$. The results received by an appropriately adapted version of SYNFOS are $\beta = 40°$ and $\Delta E = 111$ cm$^{-1}$ (Fig. 18). Actually, the second position is populated at 100 K by 17 % [62]. This quantitative result which shows that the oxygen ligand in oxy-myoglobin becomes mobile on increasing temperature is in agreement with recent studies applying DFT calculations [78]. These studies indicate qualitatively that a small (10–20 %) contribution from a second FeO$_2$ orientation will cause a major decrease in quadrupole splitting.

The temperature dependence of the jump frequency $w_{12}$ from the ground state 1 to the excited state 2 (Fig. 19) can be fitted by the Arrhenius expression

$$w(T) = w_T + w_0 \cdot e^{-E_A/k_B T} \tag{7}$$

with $w_T = 10^7$ s$^{-1}$, $w_0 = 10^9$ s$^{-1}$ and $E_A = (260\pm35)$ cm$^{-1}$. This result strongly supports the view that thermal activation plus a low-temperature tunneling process are responsible for the dynamics of the FeO$_2$ moiety in oxy-myoglobin.

Dynamical behavior of dioxygen has also been observed in model cobalt(II) hemes and cobalt-containing hybrid hemoglobin by measuring temperature-dependent EPR spectra and analysing them as a function of jump time [79]. In summary we note that the presented NFS results and also the DFT calculations for oxy-myoglobin [78] and EPR studies on $CoO_2$-hemes and -hemoglobins [79] provide new insight into the highly dynamic nature of dioxygen bound to the metal centers of myoglobin and hemoglobin at physiological temperatures.

## 5. CONCLUSIONS

NFS has been shown to permit the investigation of very small samples as, e. g., tiny single crystals, and of dynamic phenomena. Results are obtained with this method which are more precise and less ambiguous than with conventional Mössbauer spectroscopy.

NIS has been proven to be a useful vibrational spectroscopy that is complementary to IR and Raman spectroscopy. It is especially helpful as site-selective spectroscopy if, in a complex vibrational spectrum of a large molecule, only those vibrational modes in which the Mössbauer nucleus is involved are of interest. Since NIS spectra depend only on the energy and character of the vibrational modes, but not on electrical properties like polarizability, they can be interpreted more easily. Also calculations of NIS spectra by DFT methods are usually more accurate than calculations of IR or Raman intensities.

**Acknowledgement**

For helpful comments we are grateful to F.A. Walker from the University of Arizona, Tucson, presently staying with an Alexander-von-Humboldt award at the University of Lübeck.

**Figure captions**

**Fig. 1.** Scheme of the experimental setup for nuclear resonant scattering, both for NIS (Detector #1) and NFS (Detector #2).

**Fig. 2.** Experimental PDOS (●) at $T = 30$ K (LS state) for $k \perp a$ (a) and $k \parallel a$ (b). For better comparison the calculated PDOS (solid lines) is shifted and reduced in height by a factor of two (adapted from [38]).

**Fig. 3.** Experimental PDOS (●) at room temperature (HS state). For better comparison the calculated PDOS (solid line) is shifted and reduced in height by a factor of two (adapted from [38]).

**Fig. 4.** Optimized geometry of (a) the ground state, (b) the side-on bonding ($MS_2$), and (c) the isonitrosyl ($MS_1$) structure of the nitroprusside anion as retrieved from BLYP/6-311+G(2d,p) DFT calculations (adapted from [45]).

**Fig. 5.** Experimental NIS spectra at 77 K before (Δ) and after (●) warming up to 250 K for 90 minutes and cooling down again. The dashed and the dotted lines represent the simulated spectra for the ground state and $MS_1$, respectively. The solid line is a superposition of the ground state (93 %) and $MS_1$ (7 %). The peaks are labelled as described in the text (adapted from [45]).

**Fig. 6.** Temperature-dependant NFS time-domain spectra of the diamagnetic "picket-fence" porphyrin complex $^{57}FeO_2(SC_6HF_4)(TP_{piv}P)]^-$ (a). The solid lines are fits with the CONUSS program with the obtained quadrupole splitting $\Delta E_Q$ (Δ,□: data obtained from NFS. ●: data obtained from conventional Mössbauer spectroscopy) and effective thickness $t_{eff}$ plotted versus temperature in (b) and (c) (adapted from [49]).

**Fig. 7.** NFS spectra of the paramagnetic "picket-fence" porphyrin complex $^{57}Fe(CH_3COO)(TP_{piv}P)]^-$ obtained at 3.3 K in a field of 6.0 T applied (a) perpendicular to both the synchrotron beam and the polarization vector of the radiation and (b) perpendicular to the synchrotron beam but parallel to the polarization vector of the radiation. The solid lines

are simulations with the SYNFOS program using $S = 2$ and parameters described in the text (adapted from [20]).

**Fig. 8**. (a) NFS spectra of [TPPFe (NH$_2$PzH)$_2$]Cl. Spectra were recorded at 4.2 K and with applied fields as indicated. The parameters for solid lines, which represent the best fit of NFS spectra, are given in Table 2. For comparison: the dashed lines represent simulations using the parameters obtained from conventional Mössbauer spectra.
(b) Field-dependent conventional Mössbauer spectra of [TPPFe(NH$_2$PzH)$_2$]Cl. Spectra were recorded at 4.2 K and with applied fields as indicated. Dashed lines result from the simultaneous simulation of the three spectra; corresponding parameters are summarized in Table 2. For comparison: solid lines represent simulations using the parameters obtained from the NFS analysis (adapted from [52]).

**Fig. 9.** Excitation probability of Fe(TPP)(NO). The dashed curves represent the results of a fit to the excitation probability to a series of Lorentzian peaks. Lower panel: mode composition factors $e_{Fe}^2$ using the fitted peak areas from the upper panel (taken from [56] with permission of the copyright holders).

**Fig. 10.** PDOS extracted from the measured NIS spectra (solid lines) and calculated by a normal mode analysis using atomic force fields of Fe$^{II}$(TPP)(2-MeImH) for (a) in-plane (b) out-of-plane iron modes (taken from [57] with permission of the copyright holders).

**Figure 11.** Temperature-dependent NFS-spectra of deoxy-myoglobin in the temperature range from 50 K up to 243 K (taken from [60] with permission of the copyright holders).

**Fig. 12.** NIS spectrum of Mb* (photolyzed MbCO) in the region from 160 cm$^{-1}$ up to 340 cm$^{-1}$ and deconvolution of the 250 cm$^{-1}$ mode cluster of Mb* with four components with parameters given in the text. For comparison the NIS Spectrum of deoxy-myoglobin obtained at 50 K is shown below (taken from [64] with permission of the copyright holders).

**Fig. 13.** NIS spectra spectra of a met-myoglobin single crystal with its *b*-axis perpendicular to the synchrotron beam as a function of the rotation angle $\phi$ around this axis. The two met-myoglobin molecules in the unit cell together with an arrow indicating the crystallographic *b*-axis is shown in (b) (taken from [68] with permission of the copyright holders).

**Fig. 14.** Conventional Mössbauer spectra of the 'picket-fence' porphyrin complex [Fe(CH$_3$COO)(TP$_{piv}$P)]$^-$ taken at the temperatures as indicated in an external field of 4 T applied perpendicular to the γ-ray.

**Fig. 15.** Time dependence of the NFS spectra obtained from the 'picket-fence' porphyrin complex [Fe(CH$_3$COO)(TP$_{piv}$P)]$^-$ in a field of 4 T applied perpendicular to **k** and σ at the temperatures as indicated. At $T = 3$ K the NFS pattern exhibits several hyperfine frequencies due to the combined magnetic and electric quadrupole interaction, while at 30 K only a single beat frequency is visible because the magnetic hyperfine interaction has cancelled. Of special interest for the understanding of the relaxation mechanism is the transition region in between.

**Fig. 16.** Various models of the FeO$_2$ bond in oxy-myoglobin.

**Fig. 17.** NFS spectra of oxy-myoglobin at various temperatures.

**Fig. 18.** Goodness-of-the-fit $\chi^2$ and energy distance $\Delta E$ of the secondary position of the terminal oxygen as a function of the rotation angle.

**Fig. 19.** Transition rate of the terminal oxygen between the two conformations in oxy-myoglobin as a function of temperature.

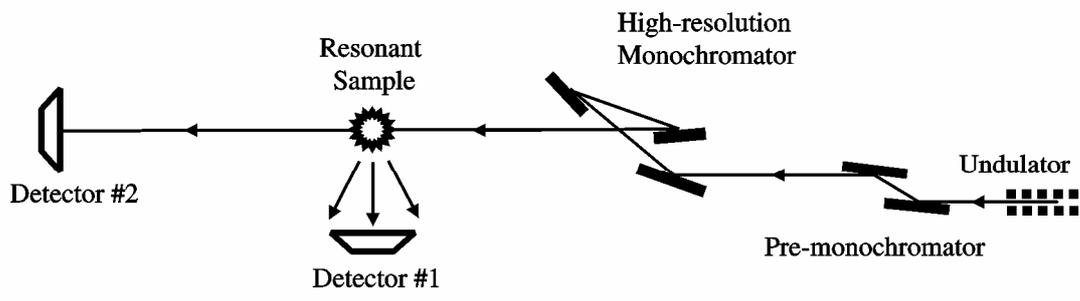

Fig. 1

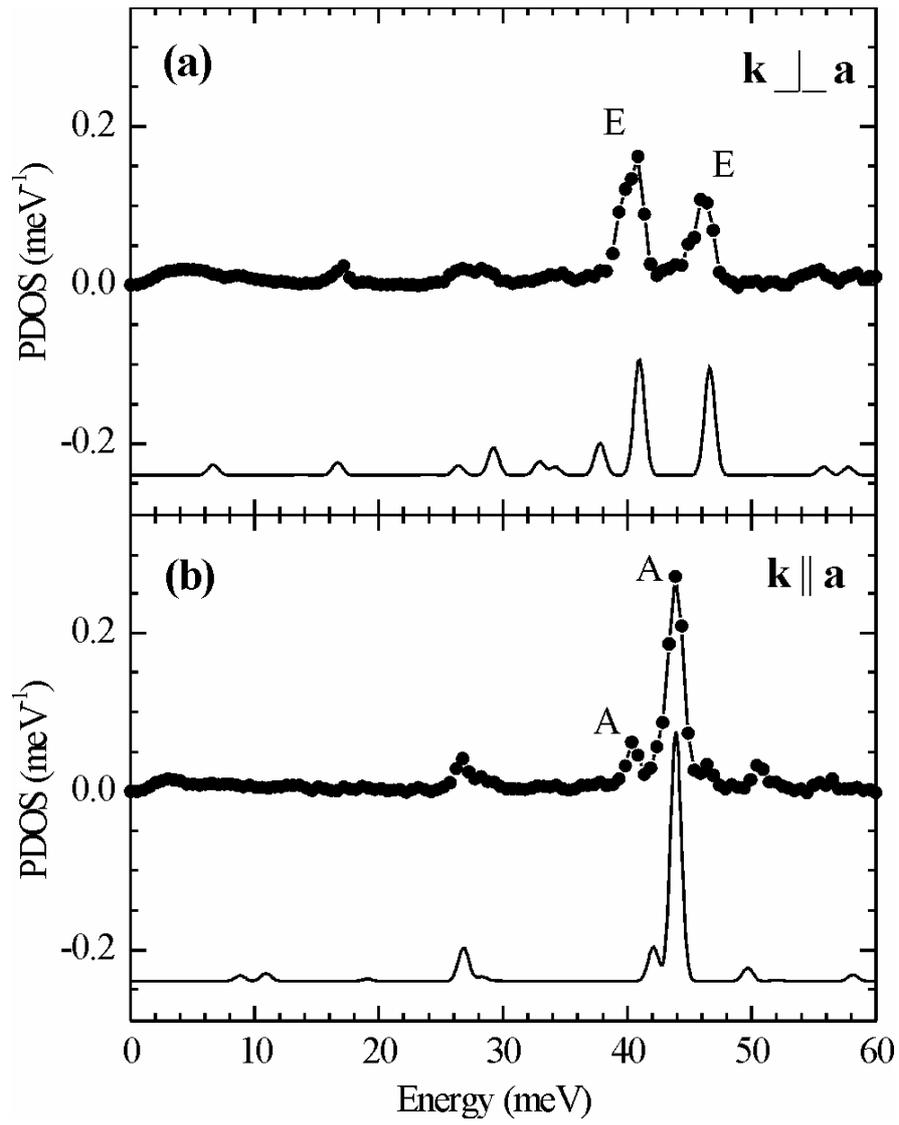

Fig. 2

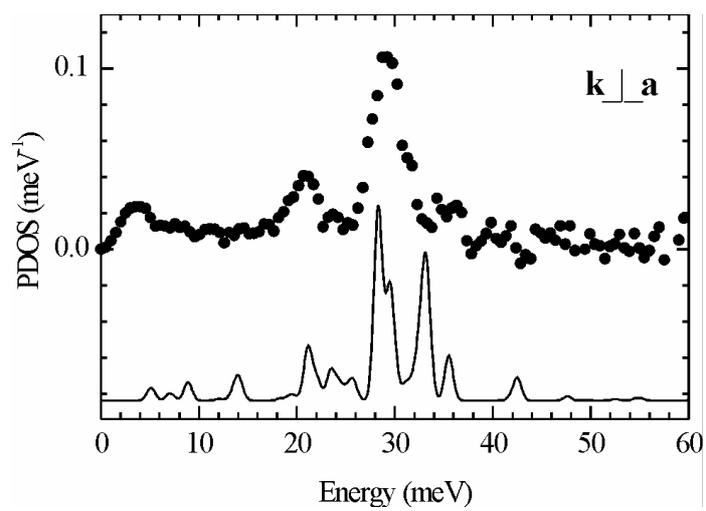

Fig. 3

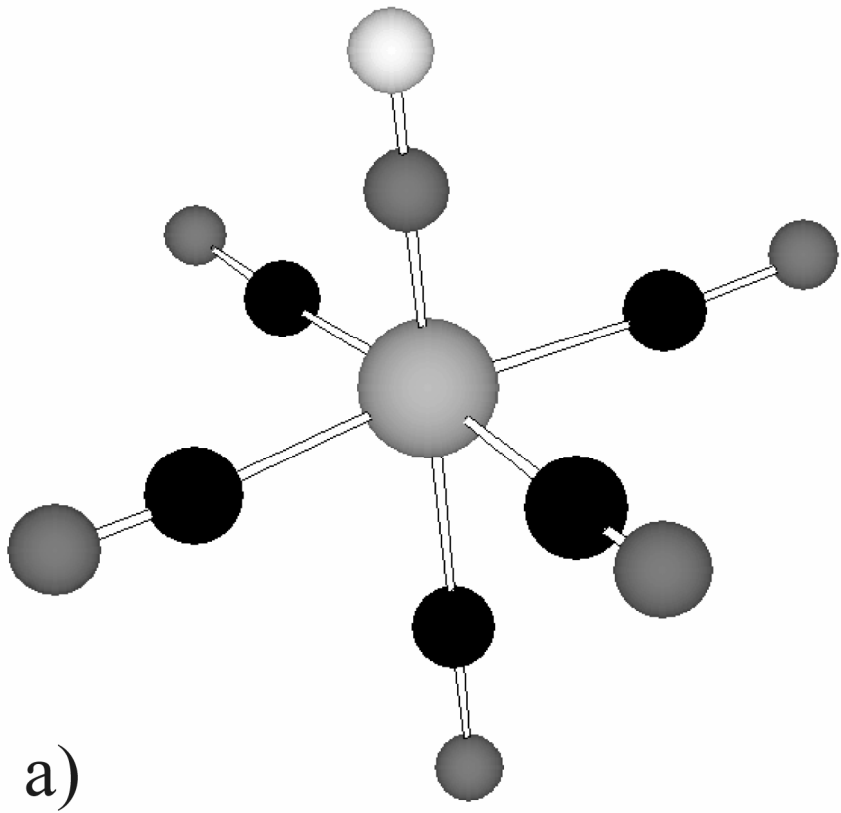

Fig. 4a

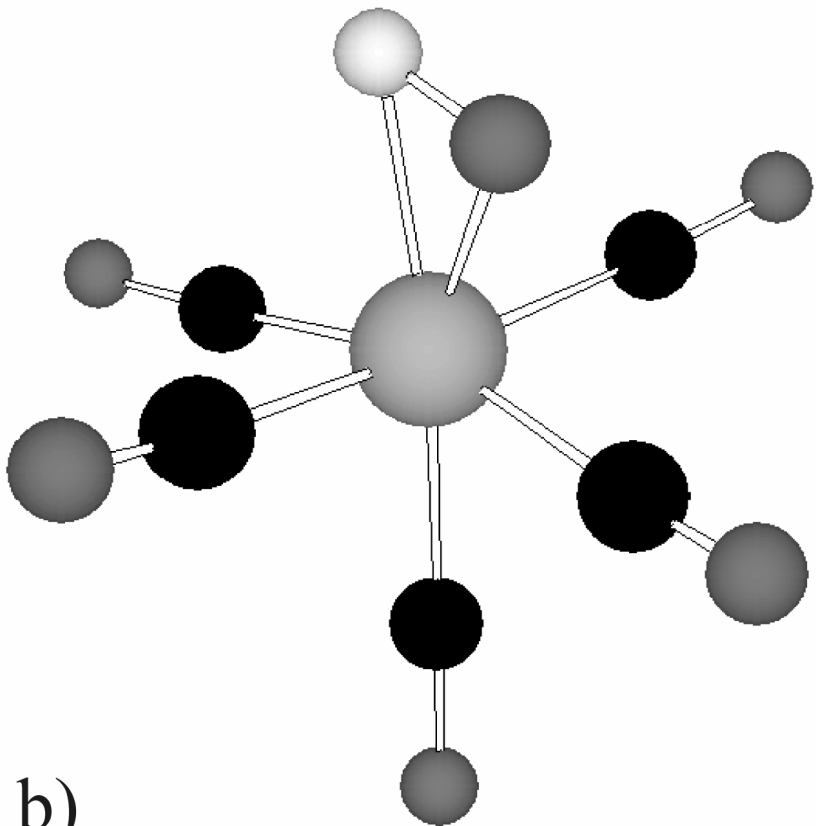

Fig. 4b

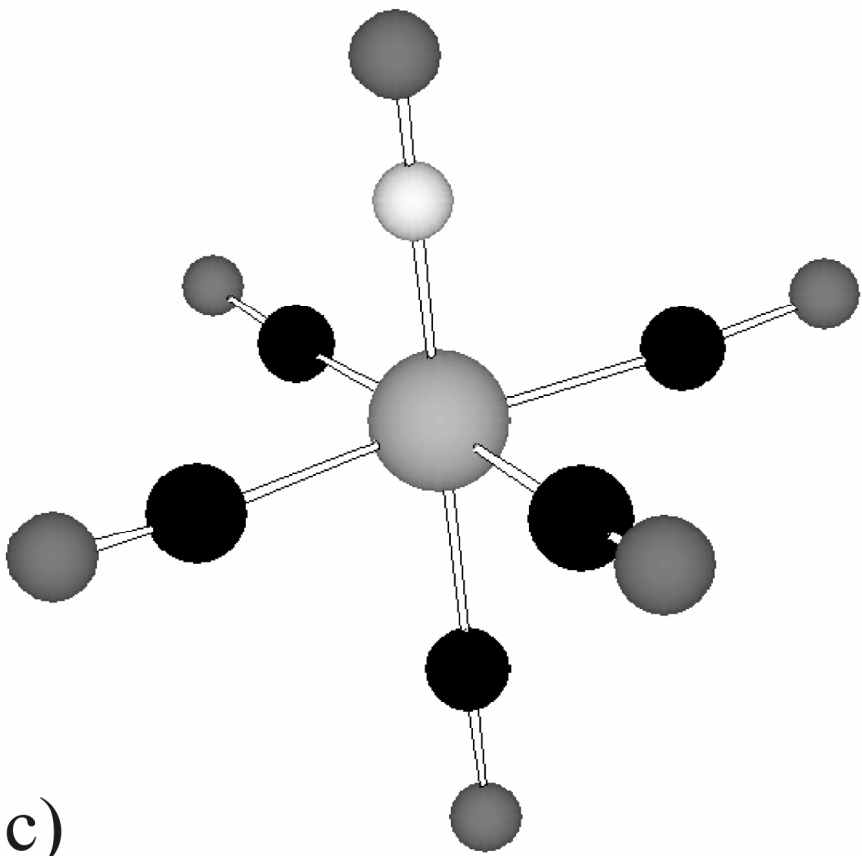

Fig. 4c

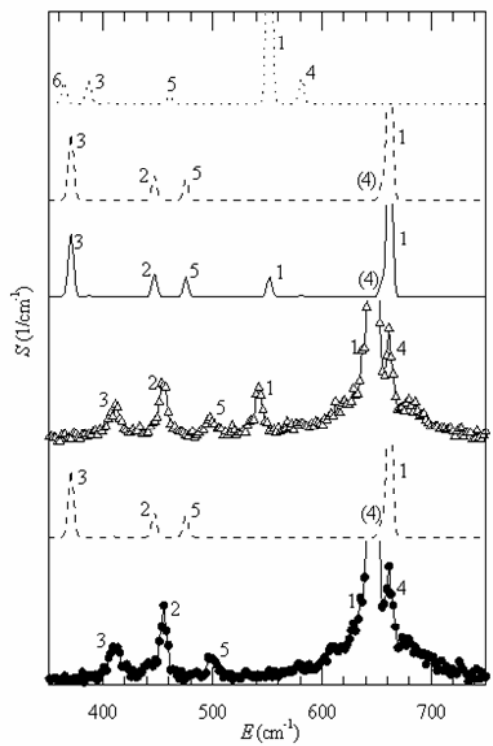

Fig. 5

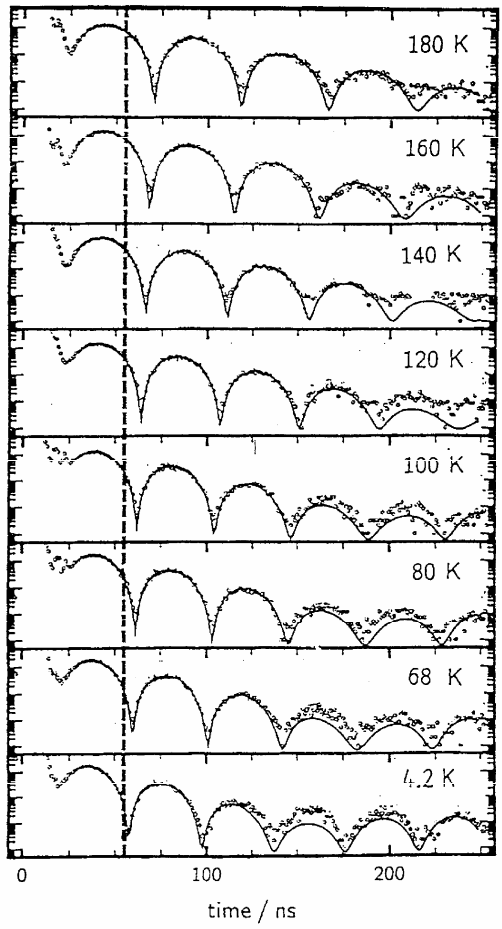

Fig. 6a

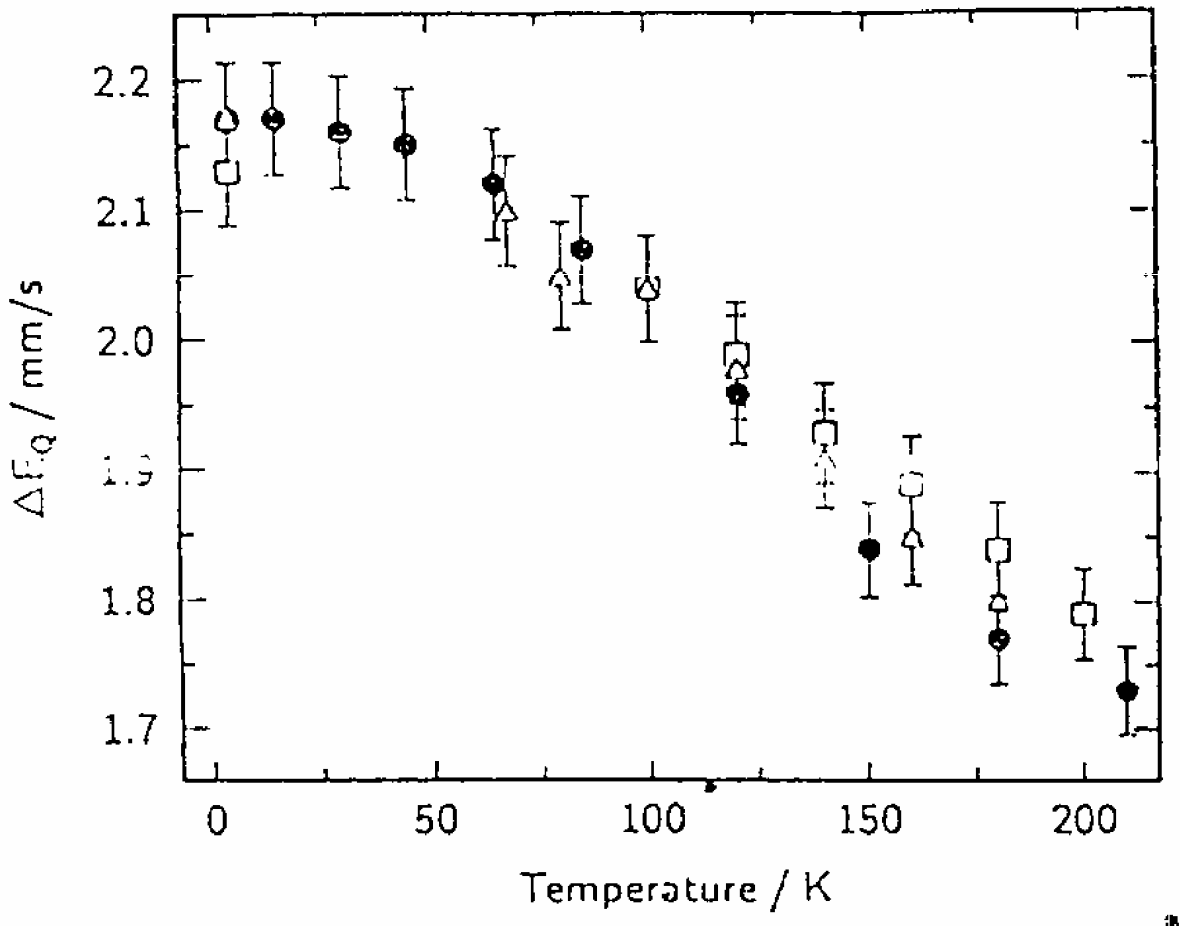

Fig. 6b

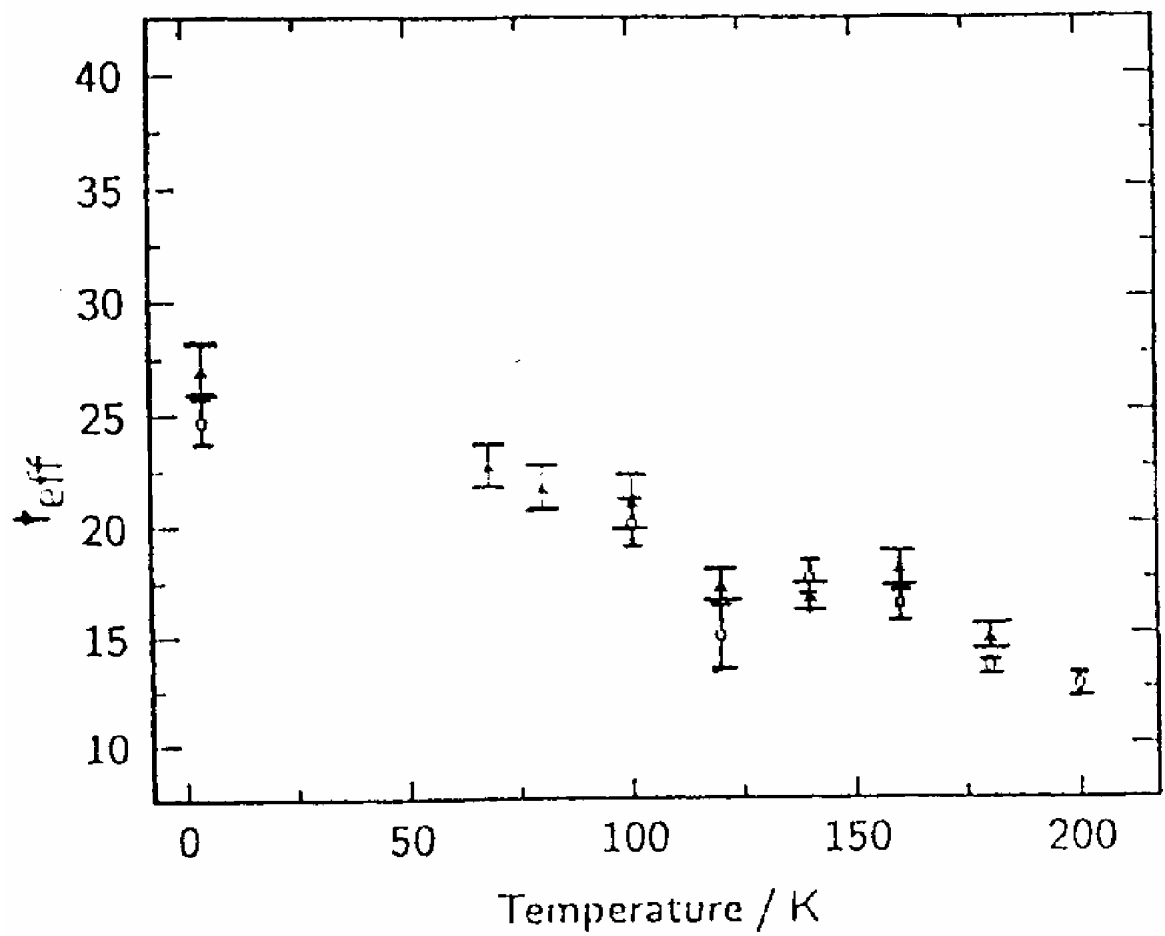

Fig. 6c

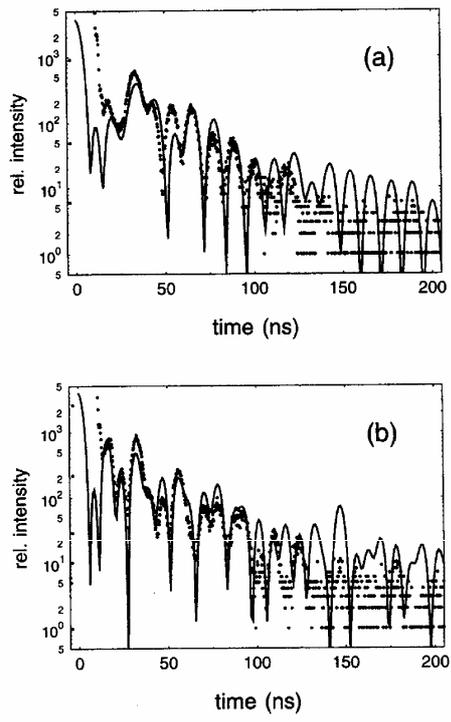

Fig. 7

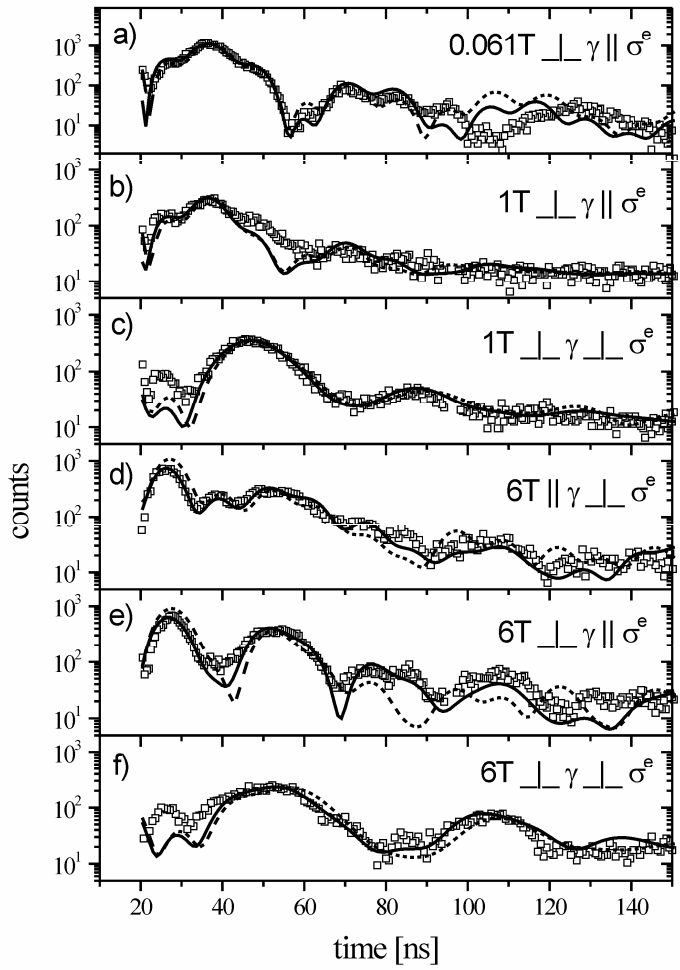

Fig. 8a

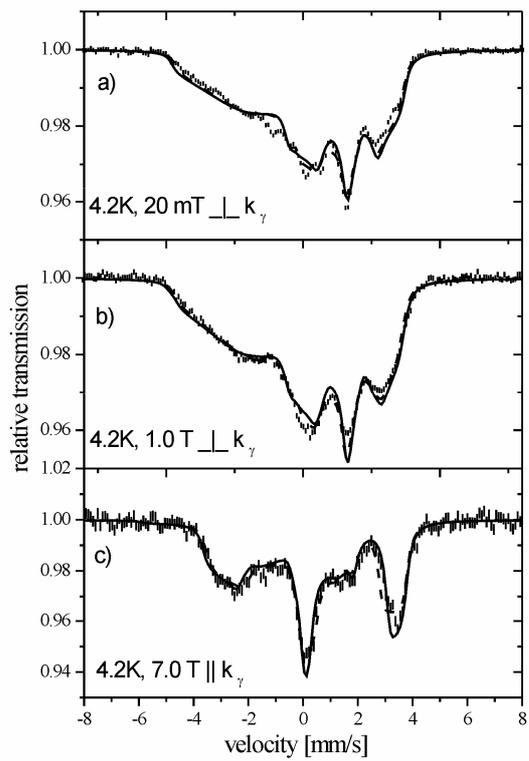

Fig. 8b

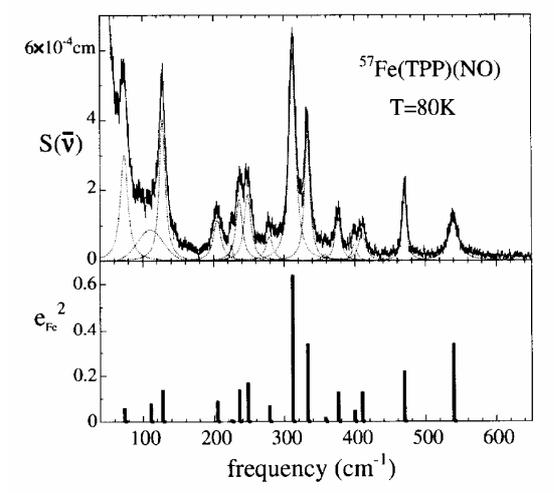

Fig. 9

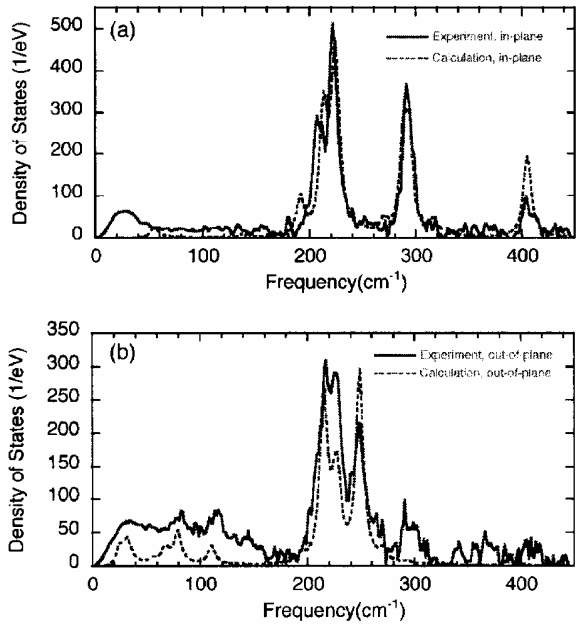

Fig. 10

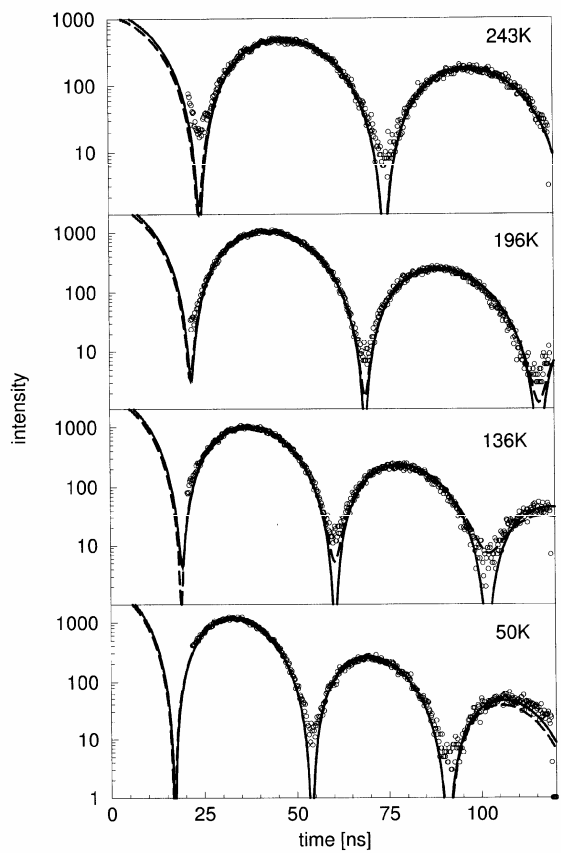

Fig. 11

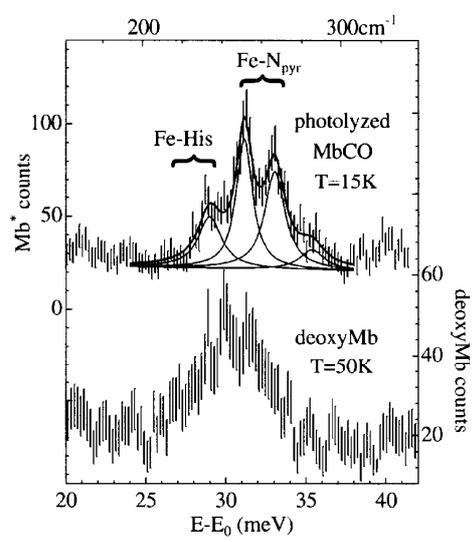

Fig. 12

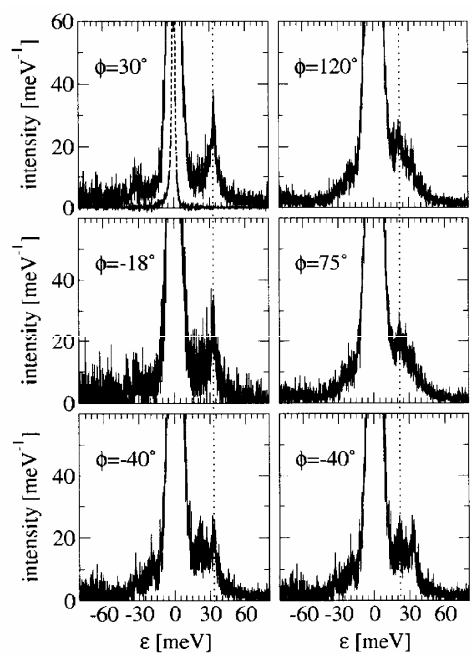

Fig. 13a

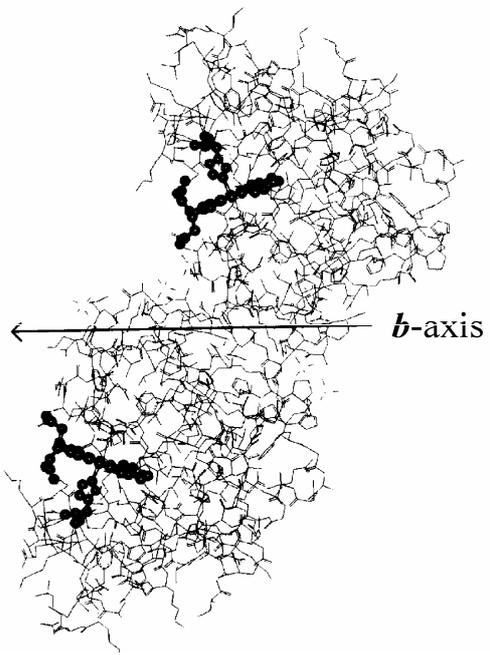

Fig. 13b

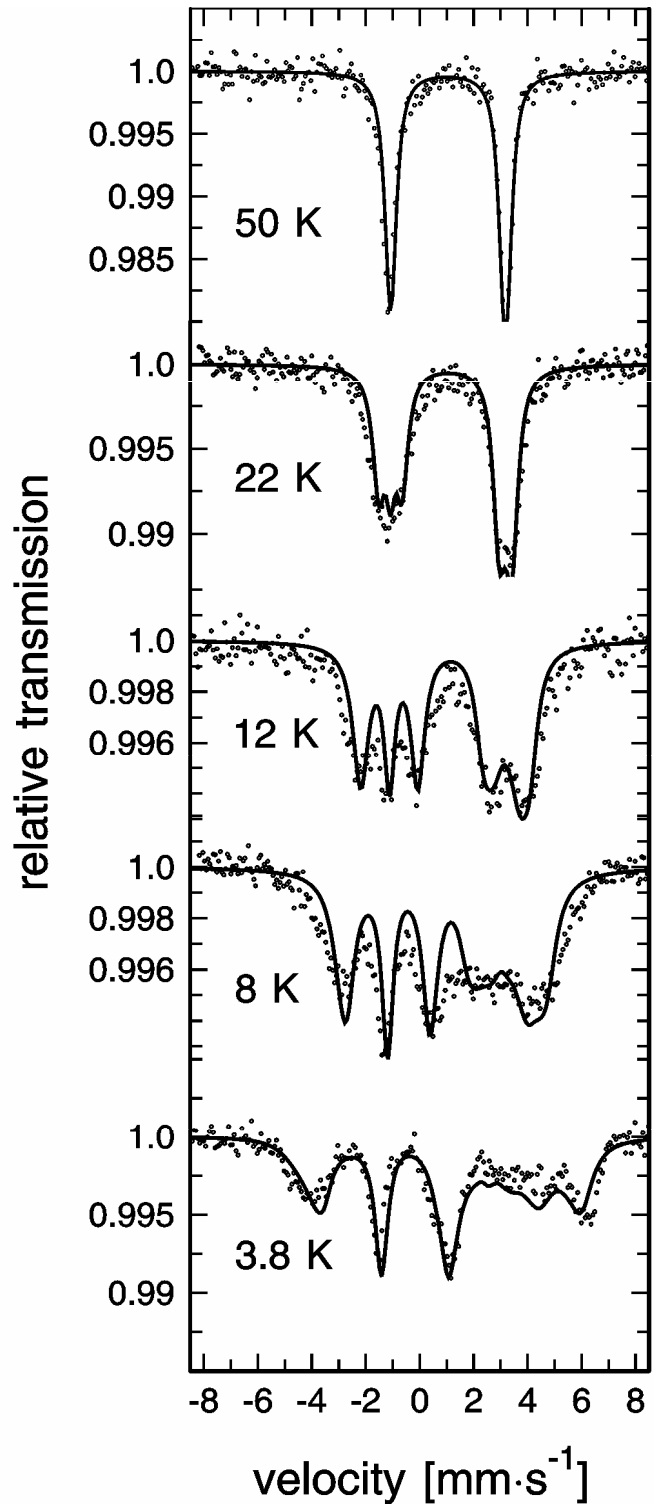

Fig. 14

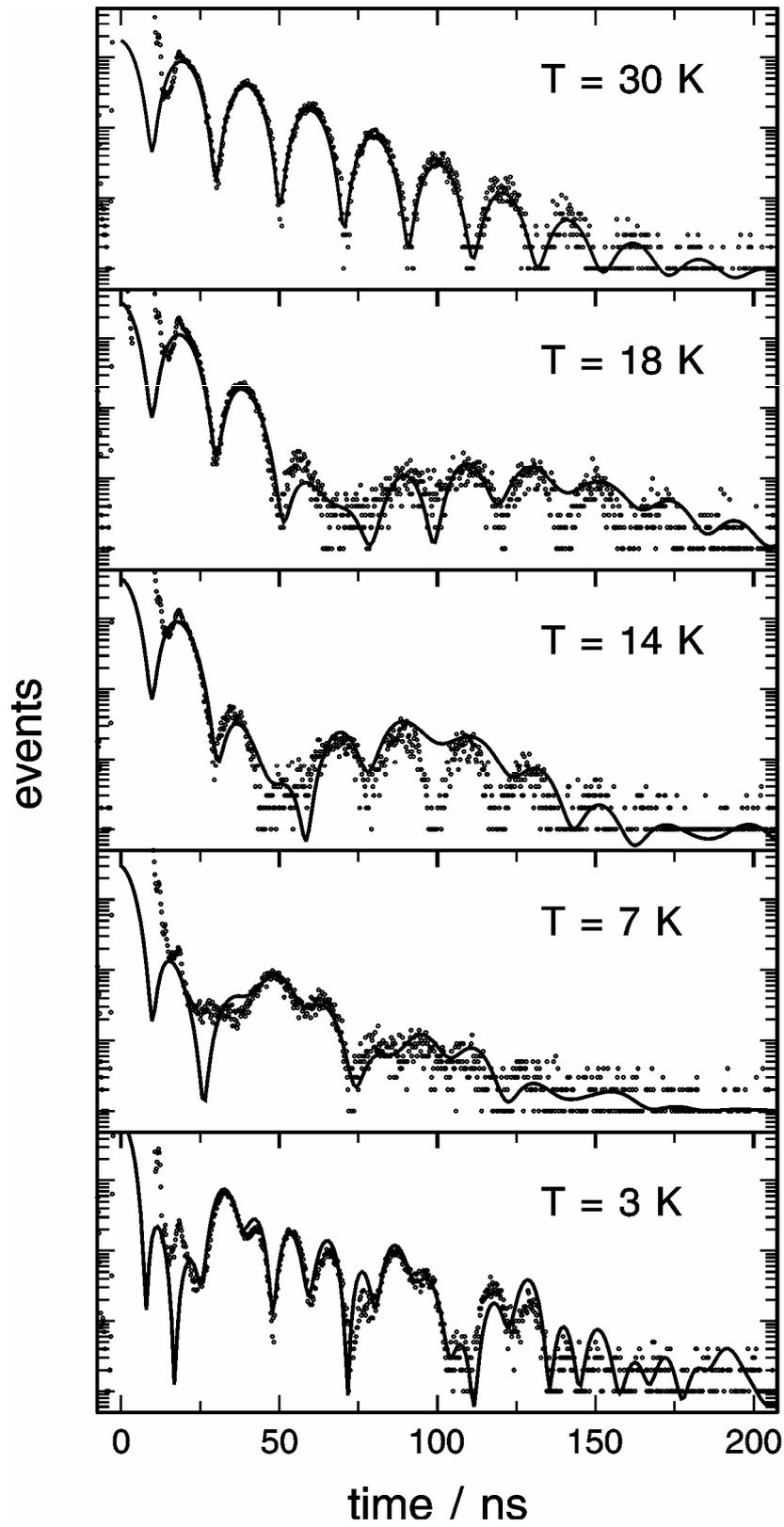

Fig. 15

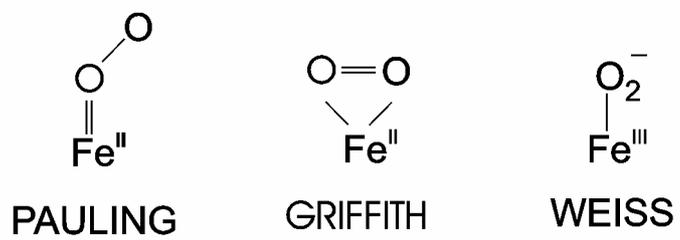

Fig. 16

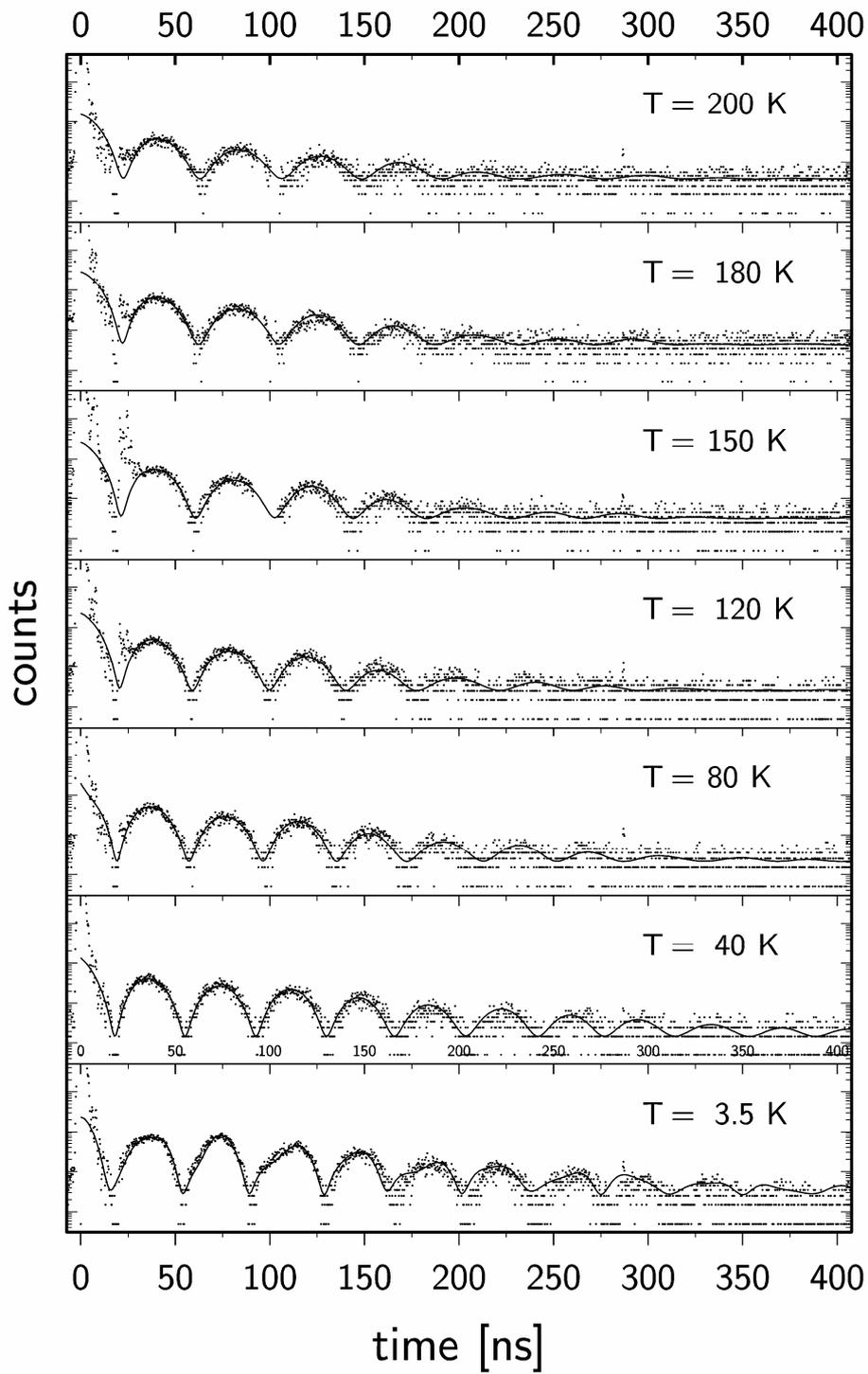

Fig. 17

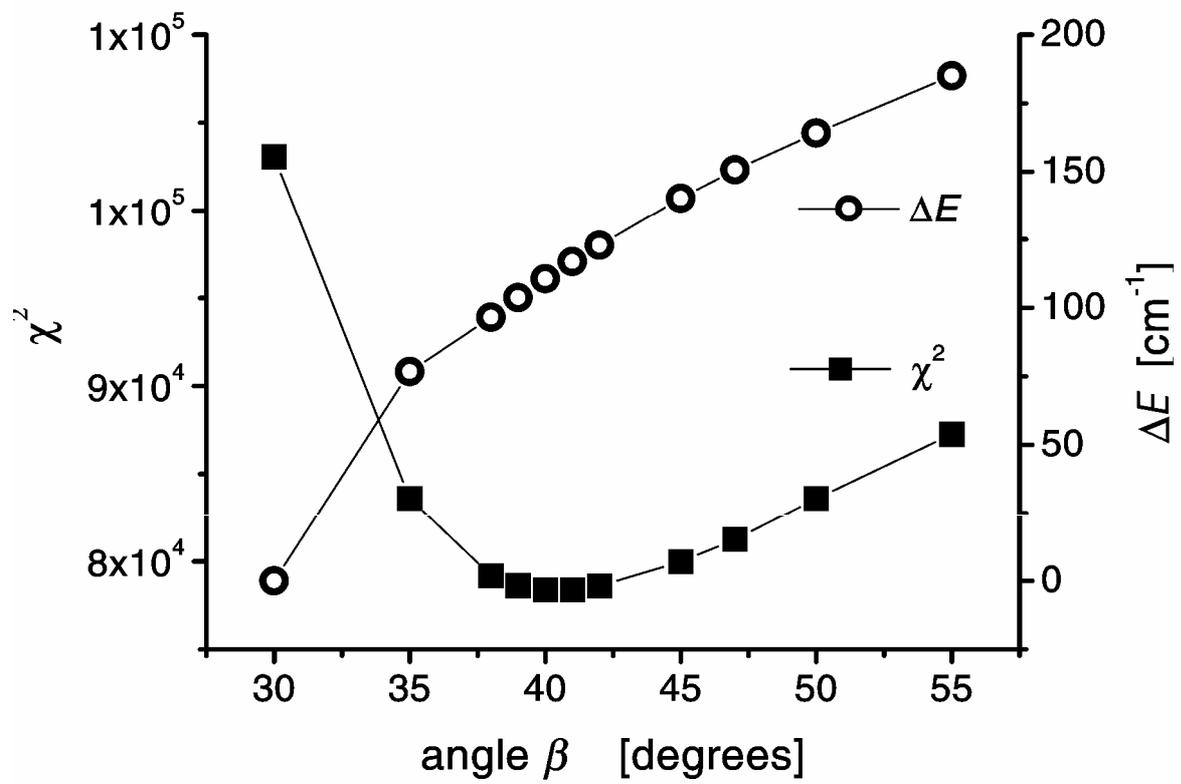

Fig. 18

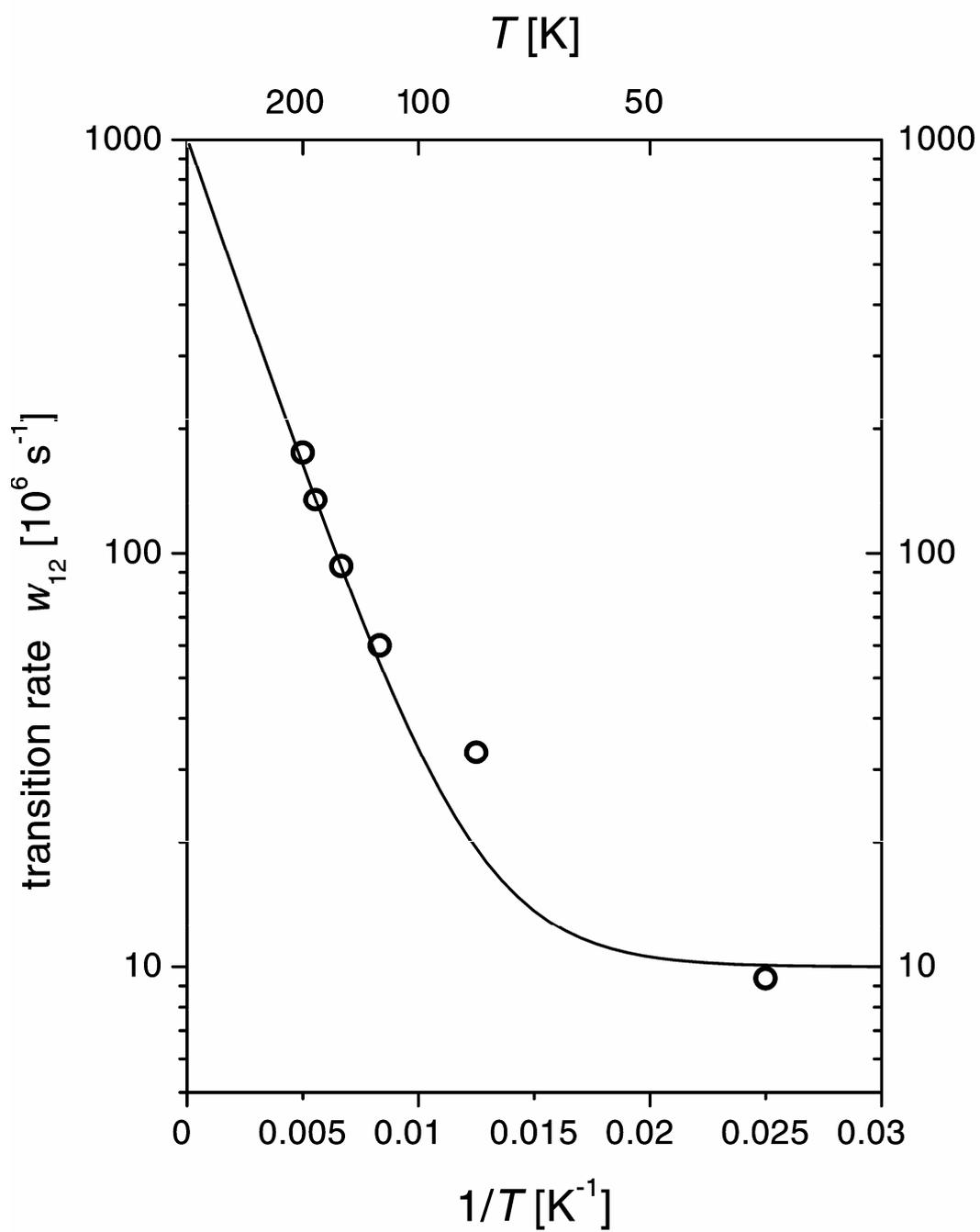

Fig. 19